\documentclass[11pt,a4paper]{article}
\usepackage[dvips]{graphicx}
\usepackage{psfrag}
\usepackage{amsmath,amssymb,amsbsy,color}
\usepackage{epstopdf}
\usepackage{grffile} % handles extra "."'s in graphics filenames
\usepackage{multirow}
\usepackage{hyperref}
\usepackage[ruled,vlined,noline]{algorithm2e}
\usepackage{algorithmic}
\usepackage[numbers,sort&compress]{natbib}
\begin{document}
\begin{flushright}
\small{
  RBRC-1235\\
  KEK-CP-358\\
}
\end{flushright}
\vspace{2mm}

\begin{center}
{\Large\bf 
Improved lattice computation of proton decay matrix elements}
\end{center}
\vspace{5mm}

\begin{center}

Yasumichi~Aoki$^{(a,b)}$,
Taku~Izubuchi$^{(b,c)}$
Eigo~Shintani$^{(d)}$, 
Amarjit~Soni$^{(c)}$
\\[4mm]
{\small\it
$^a$High Energy Accelerator Research Organization (KEK),\\
Tsukuba 305-0801, Japan}
\\
{\small\it
$^b$RIKEN-BNL Research Center, Brookhaven National Laboratory, Upton, NY 11973, USA}
\\
{\small\it
$^c$High Energy Theory Group, Brookhaven National Laboratory, Upton, NY 11973, USA}
\\
{\small\it
$^d$RIKEN Advanced Institute for Computational Science, Kobe, Hyogo 650-0047, Japan}
\\[10mm]
\end{center}

\begin{abstract}
We present an improved result of lattice computation of the proton decay matrix elements in $N_f=2+1$ QCD. In this study, the significant improvement of statistical accuracy by adopting the error reduction technique of All-mode-averaging, is achieved for relevant form factor to proton (and also neutron) decay on the gauge ensemble of $N_f=2+1$ domain-wall fermions in $m_\pi=0.34$--0.69 GeV on 2.7~fm$^3$ lattice as used in our previous work \cite{Aoki:2013yxa}. We improve total accuracy of matrix elements to 10--15\% from 30--40\% for $p\rightarrow\pi e^+$ or from 20--40\% for $p\rightarrow K \bar\nu$. The accuracy of the low energy constants $\alpha$ and $\beta$ in the leading-order baryon chiral perturbation theory (BChPT) of proton decay are also improved. The relevant form factors of $p\rightarrow \pi$ estimated through the ``direct'' lattice calculation from three-point function appear to be 1.4 times smaller than those from the ``indirect'' method using BChPT with $\alpha$ and $\beta$. It turns out that the utilization of our result will provide a factor 2--3 larger proton partial lifetime than that obtained using BChPT. We also discuss the use of these parameters in a dark matter model.  
\end{abstract}

\section{Introduction}
Although proton decay has not been observed in the experiment yet, it is
an important key observable for search of the new physics beyond the
Standard Model (SM). The observed proton lifetime, {\it i.e.}
$\tau_p>8.2\times 10^{33}$ year for $p\rightarrow \pi^0 e^+$
\cite{Nishino:2009aa} (recently $\tau_p>1.4\times 10^{34}$ year has been
reported in \cite{Babu:2013jba}) or $\tau_p>5.9\times 10^{33}$ year for
$p\rightarrow K^+\bar\nu$~\cite{Abe:2014mwa}, impose tight constraints
to the parameter space of the Grand Unified Theories (GUTs) and
supersymmetric one (SUSY-GUTs). Currently such an experimental bound
might exclude minimal SU(5) GUTs, besides SUSY-GUTs has been attractive
models for the solution of hierarchy problem and the coupling
unification of the SM in the GUT scale ($\sim 10^{16}$ GeV). SUSY-GUTs
favor $p\rightarrow K^+\bar\nu$ decay channel within the detectable
region of proton decay in the future experiment
(e.g. Hyper-Kamiokande~\cite{Kearns:2013lea}). 

%%% YA tries to make this paragraph better

The main mode of the proton decay through GUTs is those where a proton
decays into a pseudoscalar meson and an anti-lepton. The operator
product expansion (OPE) leads to the decay amplitude of such
processes written in terms of the Wilson coefficients which contain all
the details of the high energy part of a GUT, and the low energy QCD
matrix elements
of the proton and pseudoscalar states with three-quark operators. 
Each QCD matrix element is further decomposed into two
form factors, named relevant and irrelevant form factor. 
Denoting the relevant form factor $W_0$, the partial decay width reads
\begin{equation}
\Gamma(N\rightarrow P+\bar l) 
= \frac{m_N}{32\pi}\Big[ 1-\Big(\frac{m_P}{m_N}\Big)^2\Big]^2
  \Big|\sum_I C^IW_0^I(N\rightarrow P)\Big|^2 
  + O(m_l/m_N),
  \footnote{In this study we elaborate to estimate to one higher order
  in $m_l/m_N$  for
  the anti-muon final states. Simply replacing $W_0$ with $W_\mu$, which
  is described later, will provide the partial decay width which is good
  to $O(m_\mu/m_N)$.}
  \label{eq:width}
\end{equation}
with $m_N$, $m_P$ and $m_l$ being the mass of the nucleon, pseudoscalar
meson and anti-lepton, and $C^I$ being the Wilson coefficient of the
operator of type $I$ (distinguishing flavor and chiral structure),
which entering also in $W_0^I$.
Parameters in a given GUT model are encoded in the Wilson coefficients
$C^I$. The knowledge of the left hand side (experiment) and that of
$W_0$ reported in this work will be transcribed into the knowledge of
the GUT parameters. 
Namely the proton lifetime bound gives rise to restricting the
GUT parameters \cite{Martin:2011nd,Maekawa:2013sua,Maekawa:2014gva,deGouvea:2014lva,Evans:2015bxa,Mambrini:2015vna,Huo:2015nwa,Bajc:2015ita,Ellis:2015rya,Brennan:2015psa,Hisano:2015ala,Perez:2016qbo,Bajc:2016qcc,Ellis:2016tjc,Babu:2016cri,Babu:2016bmy,Kolesova:2016ibq,Cox:2016epl,Harigaya:2016vda}. 

The relevant form factors are evaluated in $\overline{\rm MS}$ scheme in the naive dimensional regularization at a typical hadronic scale $\mu=2$ GeV. The matching Wilson coefficients need to be calculated in the same way.

Lattice computation of the proton decay matrix elements has rather a long history. It is started with the calculation of low energy constants (LEC) $\alpha$ and $\beta$ in quenched approximation $N_f=0$, where one needs to use a baryon chiral perturbation theory \cite{Claudson:1981gh} to obtain $W_0$. The uncertainties of these initial computations \cite{Hara:1986hk,Bowler:1988us,Gavela:1989cp} have been successively reduced with systematic improvements by employing the direct method \cite{Aoki:1999tw,Aoki:2006ib,Aoki:2013yxa}, continuum limit of LEC's in $N_f=0$ \cite{Tsutsui:2004qc}, non-perturbative renormalization with chirally
invariant lattice formulation \cite{Aoki:2006ib,Aoki:2008ku,Aoki:2013yxa}, LEC's computed in $N_f=2+1$ \cite{Aoki:2008ku}, and finally $W_0$ computation in $N_f=2+1$ (see Table~\ref{tab:history}).

In the previous report \cite{Aoki:2013yxa} $W_0$ is calculated using the direct method with proper dynamical fermion computation using the $N_f=2+1$ domain-wall fermion formulation. This paper reported the result of $W_0$ with all the relevant systematic uncertainties removed or properly estimated. The precision, though, was not satisfactory because $W_0$ for the pion and kaon final state matrix elements have 20--40 \% errors.  Noticing the fractional error gets doubled in the partial decay width as it enters quadratically in Eq.~(\ref{eq:width}), it is necessary to have more precise results for $W_0$ in order to make them more useful. Reduction of the statistical error into sub-dominance is essential task for this study, since it has been dominated as a half and more in total error. 
%The most dominant error in that study is statistical. Therefore the next task is to reduce it, and that is what we are after in this study.

A recent development of the algorithm to speed-up the measurement of matrix element in lattice QCD, called as all-mode-averaging (AMA)~\cite{Blum:2012uh,Blum:2012my,Shintani:2014vja}, enables us to further improve the statistical precision of the proton decay matrix element for the pion and kaon final states. This paper shows the update of lattice calculation of proton decay matrix element, in both ``direct'' and ``indirect'' measurements, for all decay modes on the same gauge ensembles as used in \cite{Aoki:2013yxa}. As a consequence of increasing statistical accuracy, more reliable estimate of the systematic error can be realized.
%Although the computation here is almost following the previous one apart from the statistical accuracy, we see the systematic error of the extrapolation to physical kinematics from the simulation points is, indeed,  decreased.

In addition, previously we have not taken into account the  muon mass effect for the case of muon final state because the effect is sub-dominant compared to other uncertainty. However, with increased statistical accuracy, the effect of non-zero muon mass (106 MeV) is visible. We provide the form factors for the muon final state separately from those with positron or neutrino final states. 

Here we also attempt to use our lattice computation as an input of proton decay matrix element for the model of dark matter \cite{Davoudiasl:2010am,Davoudiasl:2011fj,Davoudiasl:2014gfa}. Although the kinematics of our setup is not optimal for those needed for the dark matter scattering there, we provide information for them as useful bi-products.

This paper is organized as follows; after showing the notation
(Section~\ref{sec:def}) and simulation parameters
(Section~\ref{sec:lat}), we show the updated result of lattice
evaluation of the low energy constants $\alpha$ and $\beta$ in BChPT 
for the indirect method in
Section~\ref{sec:lecs}, and relevant form factor $W_0$ of proton decay in
Section~\ref{sec:ff}. In Section~\ref{sec:ff} we also make an assessment
of the unestimated systematic error in the indirect method and make
alert to the use of them in the estimate of the proton lifetime.
A description of how our
results can be used in a dark matter model
\cite{Davoudiasl:2010am,Davoudiasl:2011fj,Davoudiasl:2014gfa} is given
in Section~\ref{sec:IND}. A test of the soft-pion theorem of our lattice
results is shown in Appendix~\ref{appx:soft}. Throughout the paper
dimension-full quantities are expressed in the lattice unit and the
lattice spacing ``$a$'' is suppressed in equations.

\section{Proton decay matrix element}\label{sec:def}
Lattice calculation is concentrating on the QCD matrix element of
$N\rightarrow P$ transition, in which $N$ denotes the nucleon (proton or
neutron) , and $P$ is one of pseudoscalar from $\pi^0$, $\pi^\pm$,
$K^0$, $K^+$ or $\eta$ mesons. At the hadronic energy scale, only lowest
dimensional operators with baryon number violation are relevant. They are 
the dimension-six four-fermi (three quarks and one lepton) operators
\cite{Weinberg:1979sa,Wilczek:1979hc,Abbott:1980zj}. Since the on-shell
lepton can be omitted from QCD matrix element, the transition form
factor from a nucleon $N(k)$ state (source field) to meson $P(p)$ state
(sink field) with a momentum transfer $q=k-p$ is represented as 
\begin{equation}
  \langle P(p)|\mathcal O^{\Gamma\Gamma'}(q)|N(k,s)\rangle
  = P_{\Gamma'}\Big[ W_0^{\Gamma\Gamma'}(q^2) - \frac{iq\hspace{-2mm}/}{m_N}W_1^{\Gamma\Gamma'}(q^2)\Big]
    u_N(k,s).
    \label{eq:formfactor}
\end{equation} 
On the physical kinematics $-q^2=m_l^2$, the contribution of the $W_1$ term is relatively small compared to the $W_0$ term, because for the suppression prefactor $m_l/m_N$. In $m_l=e^+$ or $\bar\nu$, $W_0$ is only relevant to proton decay matrix element since the second term is $m_{e}/m_N \sim \mathcal O(10^{-3})$, while in $m_l=m_\mu$ case, because of $m_\mu/m_N \sim \mathcal O(10^{-1})$, the contribution of $W_1$ term to the matrix element is not negligible for our target precision (below 10\% precision), namely we define
\begin{equation}
  W_\mu\equiv W_0(-m_\mu^2) + \frac{m_\mu}{m_N}W_1(-m_\mu^2).
  \label{eq:w_mu}
\end{equation}
We estimate $W_0$ and $W_1$ and provide $W_0(q^2=0)$ for the positron
and neutrino final states and $W_\mu$ for the anti-muon final states.
Baryon number violating three-quark operator $O^{\Gamma\Gamma'}$ reads
\begin{equation}
  O^{\Gamma\Gamma'} = U_{\Gamma\Gamma'}\varepsilon^{ijk}(q^{i\,T}CP_\Gamma q^j)P_{\Gamma'}q^k,
  \label{eq:BVOp}
\end{equation}
with chiral projection $P_\Gamma = (1\pm\gamma_5)/2$, where ``+'' is for $\Gamma=R$ and ``$-$`` is for $\Gamma=L$. $q^i$ is quark flavor of up, down and strange with color index $i$. $U_{\Gamma\Gamma'}$ denotes the renormalization factor which has been already computed by non-perturbative method \cite{Aoki:2008ku}
\footnote{Recently we found an error in our one-loop perturbative formula which is used for matching between $\overline{\rm MS}$ with naive dimensional regularization and RI-SMOM renormalization scheme, Eq.~(C.8) in \cite{Aoki:2006ib} and Eq.~(46) in \cite{Aoki:2008ku} (we thank Michael Buchoff and Michael Wagman for pointing out that mistake). After correction, the impact of all matrix element calculations is about 6--7\% increase for $\alpha$, $\beta$ and $W_0$ \cite{Aoki:2006ib,Aoki:2008ku,Aoki:2010dy}. In this paper we use the corrected value as presented in Eq.~(\ref{eq:zfact}).}. Using the symmetry of parity transformation between different chirality combinations $RL\Leftrightarrow LR$ or $LL\Leftrightarrow RR$ \cite{Aoki:2006ib} enables us to reduce four chirality combinations to two combinations, $\Gamma\Gamma'=LL$ and $RL$. Applying the exchange symmetry of $u$ and $d$, we have the equivalence of matrix elements between proton and neutron, 
\begin{eqnarray}
\langle \pi^0|(ud)_{\Gamma}u_{\Gamma'}|p\rangle &=& \langle \pi^0|(du)_{\Gamma}d_{\Gamma'}|n\rangle,\\
\langle \pi^+|(ud)_{\Gamma}d_{\Gamma'}|p\rangle &=& -\langle \pi^-|(du)_{\Gamma}u_{\Gamma'}|n\rangle,\\
\langle K^0|(us)_{\Gamma}u_{\Gamma'}|p\rangle &=& -\langle K^+|(ds)_{\Gamma}d_{\Gamma'}|n\rangle,\\ 
\langle K^+|(us)_{\Gamma}d_{\Gamma'}|p\rangle &=& -\langle K^0|(ds)_{\Gamma}u_{\Gamma'}|n\rangle,\\ 
\langle K^+|(ud)_{\Gamma}s_{\Gamma'}|p\rangle &=& -\langle K^0|(du)_{\Gamma}s_{\Gamma'}|n\rangle,\\ 
\langle K^+|(ds)_{\Gamma}u_{\Gamma'}|p\rangle &=& -\langle K^0|(us)_{\Gamma}d_{\Gamma'}|n\rangle,\\ 
\langle \eta|(ud)_{\Gamma}u_{\Gamma'}|p\rangle &=& -\langle \eta|(du)_{\Gamma}d_{\Gamma'}|n\rangle, 
\end{eqnarray}
and furthermore, for $p\rightarrow \pi$ channel, there is a relation in
the SU(2) isospin limit, which is good for our target precision,
\begin{equation}
  \langle \pi^+|(ud)_{\Gamma}d_{\Gamma'}|p\rangle = \sqrt{2}\langle \pi^0|(ud)_{\Gamma}u_{\Gamma'}|p\rangle,
\end{equation}
and therefore the total number of matrix element ends up to twelve. In this paper we show twelve principal matrix elements of $\langle P|\mathcal O^{\Gamma L}|p\rangle$, for $\Gamma=R$ and $L$ in lattice QCD.

Our target matrix element can be extracted from the computation of ratio given from three-point function and two-point function. We use the same combination used in Eq.~(21) of \cite{Aoki:2010dy},
\begin{eqnarray}
  R^{\Gamma L}_3(t,t_1,t_0;\vec p;P) &=&
  \frac{{\rm tr}[PC_{\mathcal O^{\Gamma L}}(t_1,t,t_0;\vec p)]}
       {C_P(t_1,t;\vec p){\rm tr}[P_4C_N(t,t_0)]} \sqrt{Z_P Z_N},
    \label{eq:ratio}
\end{eqnarray}
with nucleon two-point function $C_N(t,t_0)$ without momentum, and pseudoscalar two-point function $C_P(t,t_0;p)$ with spatial momentum $\vec p$. Here we also use the two projection matrices, $P=P_4\equiv (1+\gamma_4)/2$ and $iP_4\gamma_j$. Three-point function $C_{\mathcal O^{\Gamma L}}(t_1,t,t_0;p)$ depends on $t-t_0$, the time-slice of operator, and, $t_s=t_1-t_0$, source-sink separation, and also injected momentum $\vec p$ in the operator. The factors $\sqrt{Z_P}$ and $\sqrt{Z_N}$ are overlap factors of the pseudoscalar and nucleon states to their interpolating operators. Asymptotic form of this ratio taking the trace with two projection matrices $P_4$ and $iP_4\gamma_j$ can be expressed as the combination of $W_0$ and $W_1$
\begin{eqnarray}
  \lim_{t_1-t,t-t_0\rightarrow \infty} R^{\Gamma L}_3(t,t_1,t_0;\vec p,P_4)
  &=& W_0^{\Gamma L}+ \frac{m_N-E_\pi}{m_N}W_1^{\Gamma L},
  \label{eq:proj0}\\
  \lim_{t_1-t,t-t_0\rightarrow \infty} R^{\Gamma L}_3(t,t_1,t_0;\vec p,iP_4\gamma_j)
  &=& \frac{q_j}{m_N}W_1^{\Gamma L},
  \label{eq:projj}
\end{eqnarray}
and solving the linear algebra we derive $W_0^{\Gamma L}$ and $W_1^{\Gamma L}$ simultaneously.

Calculating the three-point function $C_{\mathcal O^{\Gamma
L}}(t_1,t,t_0;p)$ in Eq.~(\ref{eq:projj}) involves several steps. First
we compute the forward quark propagator with the nucleon source located
at $t=t_0$ with a smeared source. Then using the propagator at the meson
sink position $t=t_1$ the sequential source computation is applied with
an injection of momentum $\vec{p}$. Then the obtained backward
propagator is contracted at the operator position $t=t$ with two forward
propagators from the nucleon source. This process needs $1+N_p\times 2$
solver computations for each gauge configuration, with $N_p$ being the
number of different meson momenta. ``1'' is for the forward propagator
and the distinction of valence mass for the $ud$ and $s$ quarks makes
the factor ``2''. For the good constraint on the fitting parameters, we
need to have good lever arm for $m$ (different ensembles) and
variation of $\vec{p}$, which tends to sum up a large computational
cost. The all-mode-averaging (AMA)
technique~\cite{Blum:2012uh,Blum:2012my,Shintani:2014vja,vonHippel:2016wid} is useful to
reduce the computational cost of quark propagator by using this
method. It enables us to carry out the high statistical measurement even
using several momenta.

\section{Lattice setup}\label{sec:lat}
We use the same lattice gauge ensembles as used in \cite{Aoki:2010dy,Arthur:2012opa,Blum:2014tka}, which are generated with $N_f=2+1$ the domain-wall fermion (DWF) and Iwasaki gauge action at $\beta=2.13$, corresponding to $a^{-1}=1.7848(6)$ GeV~\cite{Blum:2014tka}, in 24$^3\times$64 lattice size ($\simeq2.65$ fm$^3$). The four different quark masses, $m=0.005$, 0.01, 0.02 and 0.03 are used in unitary point and the corresponding pion, kaon and nucleon masses are given in Table \ref{tab:latparam}. In the measurement of two-point function of pseudoscalar meson and nucleon, we use the gauge invariant Gaussian smeared source and sink function with interpolation operator on APE smeared link variable, whose parameter is same as \cite{Aoki:2010dy}. For ``indirect'' method, two-point function including baryon number violating operator Eq.(\ref{eq:BVOp}) is computed using two nucleon source operators 
\begin{eqnarray}
\mathcal N_5 &=& \varepsilon^{ijk}(u^{iT}C\gamma_5d^j)u^k,\\
\mathcal N_{45} &=& \varepsilon^{ijk}(u^{iT}C\gamma_4\gamma_5d^j)u^k,
\end{eqnarray}
and averaged them. On the other hand, in ``direct'' method for the
two-point function in the ratio in Eq.~(\ref{eq:ratio}), we use only
$\mathcal N_5$, as proton interpolation operator.
%since the former study
%\cite{Aoki:2013yxa} indicates that there is strong correlation between
%two operators in three-point function. 

The renormalization factor to make the operators to ones in $\overline{\rm MS}$ naive dimensional regularization (NDR) scheme at $\mu=2$ GeV is calculated by RI/MOM non-perturbative renormalization method combined with the RI/MOM $\to$ $\overline{\rm MS}$ matching factor calculated to the next-to-leading order in perturbation theory.
% The updated 
%
%\footnote{Those reported in Ref.~\cite{Aoki:2008ku} uses the matching factor in Ref.~\cite{Aoki:2006ib}, which turned out to have an error. Here we give the update on the factors from Ref.~\cite{Aoki:2008ku} with the error corrected. Now the values of these factors are 6-7 \% larger than the original.}
%
The renormalization factors of $\mathcal O^{\Gamma}$ with $\overline{\rm MS}$ NDR at $\mu=2$ GeV are given as 
\begin{equation}
  U^{RL} = 0.705(11)(56),\quad U^{LL} = 0.706(11)(56), 
  \label{eq:zfact}
\end{equation}
where the first error is statistical, and second is systematic. The systematic error is dominated by the truncation error in perturbative matching, which is done to next-to-leading order. The estimate is from the size of $\alpha_s^2(\mu = 2 \mbox{GeV})$ obtained by the RGE running starting from $\alpha_s(M_Z)=0.1176(2)$ (see \cite{Aoki:2006ib}). 

In the computation of three-point function, we use $t_s=18$ ($\simeq 1.98$ fm) for source-sink separation which is shorter than $t_s=22$ ($\simeq 2.43$ fm) used in \cite{Aoki:2013yxa}. Although the usage of short source-sink separation will make a suppression of the statistical noise, we need to make sure the excited state contamination is negligible. As the contamination would be more serious for smaller quark mass, we test such a contamination effect by comparing the ratio $R_3^{\Gamma L}$ for $t_s=18$ and $22$ at the lightest quark mass in Section \ref{sec:lecs}.

We use three non-zero spatial momenta for the mesons: $\vec p=(1,0,0)$, (1,1,0) and (1,1,1), where the last one is a new addition from the previous study \cite{Aoki:2013yxa}. This will provide a good lever arm for the $p^2$ direction as well as the good reach for the momentum range in the different kinematics (see Section~\ref{sec:IND}).
%\cite{Davoudiasl:2010am,Davoudiasl:2011fj,Davoudiasl:2014gfa}.

The AMA technique is applied to the measurement of the three-point and two-point functions \footnote{For the two-point function $C_P(t_1,t;\vec p)$ of pion (or eta) in the denominator of Eq.~(\ref{eq:ratio}) we have used Kuramashi-wall source as in \cite{Aoki:2010dy} for heavier ``light'' quark mass, $m=0.02$ and $m=0.03$, since there is less gain for cost per precision of signal, and thus AMA was not applied in this case.}. The low-mode deflation is used in solving the even-odd decomposed Dirac kernel with conjugate gradient method. Corresponding low-mode is computed by Lanczos algorithm with Chebyshev polynomial acceleration as performed in \cite{Shintani:2014vja}. The number of low-modes $N_{\rm eig}$ we computed for each quark mass are given in Table~\ref{tab:latparam}. Approximation used in AMA is also constructed by the sloppy solver using relaxed stopping condition (0.003 for the squared norm, which is compared with the $10^{-8}$ for the ``exact solve'' done once every configuration). $N_g$ shown in Table~\ref{tab:latparam} presents the number of such an approximation we use in AMA. Note that in the strange quark mass we use sloppy solver without deflation to avoid the additional computation of low-mode. Even without low-mode of strange quark, AMA is also working well. Actually we check that correlation between exact and approximation is smaller than $1/N_g$. 

\begin{table}
\begin{center}
\caption{Lattice ensemble set and parameters. $m$ refers to the domain-wall fermion mass for the degenerate light quarks ($u$ and $d$).  $N_g$ is the number of approximate in both light and strange quark propagators. $N_{\rm eig}$ denotes the number of low-mode used in light quark propagator, and ``res'' is value of squared norm of residual vector for the sloppy solver in AMA. Values of hadron masses are measured with extended quark source and sink with gauge invariant Gaussian smearing.
%In the row of $m=0.005$ the value marked as $^*$ is in $t_s=22$ case, and others are in $t_s=18$. 
}
\label{tab:latparam}
\begin{tabular}{ccccccccc}
\hline\hline
$a^{-1}$ GeV & $m$ & $m_\pi$ (GeV) & $m_K$ (GeV) & $m_N$ (GeV) & $N_{\rm eig}$ & $N_g$ 
& res & $N_{\rm conf}$\\
\hline
1.7848(6)
& 0.005 & 0.340(1) & 0.594(2) & 1.179(5) & 300 & 32 & 0.003 & 91\\
& 0.01 & 0.427(1) & 0.626(1) &  1.269(5) & 300 & 32 & 0.003 & 55 \\
& 0.02 & 0.574(1) & 0.688(1) & 1.452(4) & 200 & 32 & 0.003 & 39 \\
& 0.03 & 0.694(1) & 0.744(2) & 1.598(5) & 200 & 32 & 0.003 & 44 \\
\hline\hline
\end{tabular}
\end{center}
\end{table}

\section{Improved result of low energy constants}\label{sec:lecs}
First we update the ``indirect'' measurement from computation of low energy constants (LECs) for baryon number violating interaction in chiral Lagrangian~\cite{Claudson:1981gh} following the method in~\cite{Aoki:1999tw,Tsutsui:2004qc,Aoki:2006ib,Aoki:2008ku,Braun:2008ur}. In the ``indirect'' measurement, once corresponding LECs are obtained by lattice QCD, through baryon chiral perturbation theory (BChPT) together with nucleon mass, couplings to axial current (axial charge), pion decay constant and its mass, the proton decay amplitude can be evaluated. Each matrix element is proportional to LECs depending on chirality; $\alpha$ (for $RL$) and $\beta$ (for $LL$) \cite{Claudson:1981gh,Aoki:1999tw}. Those are defined through the nucleon to vacuum matrix elements. Writing the quark flavor explicitly 
\begin{equation}
  \langle 0 | (ud)_R u_L|p\rangle = \alpha P_Ru_p,\quad
  \langle 0 | (ud)_L u_L|p\rangle = \beta  P_Lu_p,
\end{equation}
with proton spinor field $u_p$. The above matrix elements can be extracted from the ratio of two-point function at large time-slice separation, 
\begin{equation}
  R_\alpha = \frac{C^R_{N\mathcal O}(t)}{C_N(t)}Z_N \xrightarrow{t\rightarrow\infty} \alpha,\quad 
  R_\beta = \frac{C^L_{N\mathcal O}(t)}{C_N(t)}Z_N \xrightarrow{t\rightarrow\infty} \beta
\end{equation}
with the nucleon decay operator ${\mathcal O}$ and the nucleon interpolating operator having the same flavor content. The nucleon overlap factor $Z_N$ is also calculated from the nucleon two-point function. From the practical point of view, this method is much cheaper than the ``direct'' method, since $\alpha$ and $\beta$ are obtained with single computation of quark propagator at each quark mass. Whereas, the direct method needs at least additional two propagators for each momentum values in the computation of three-point function.

\begin{figure}
\begin{center}
\includegraphics[width=140mm]{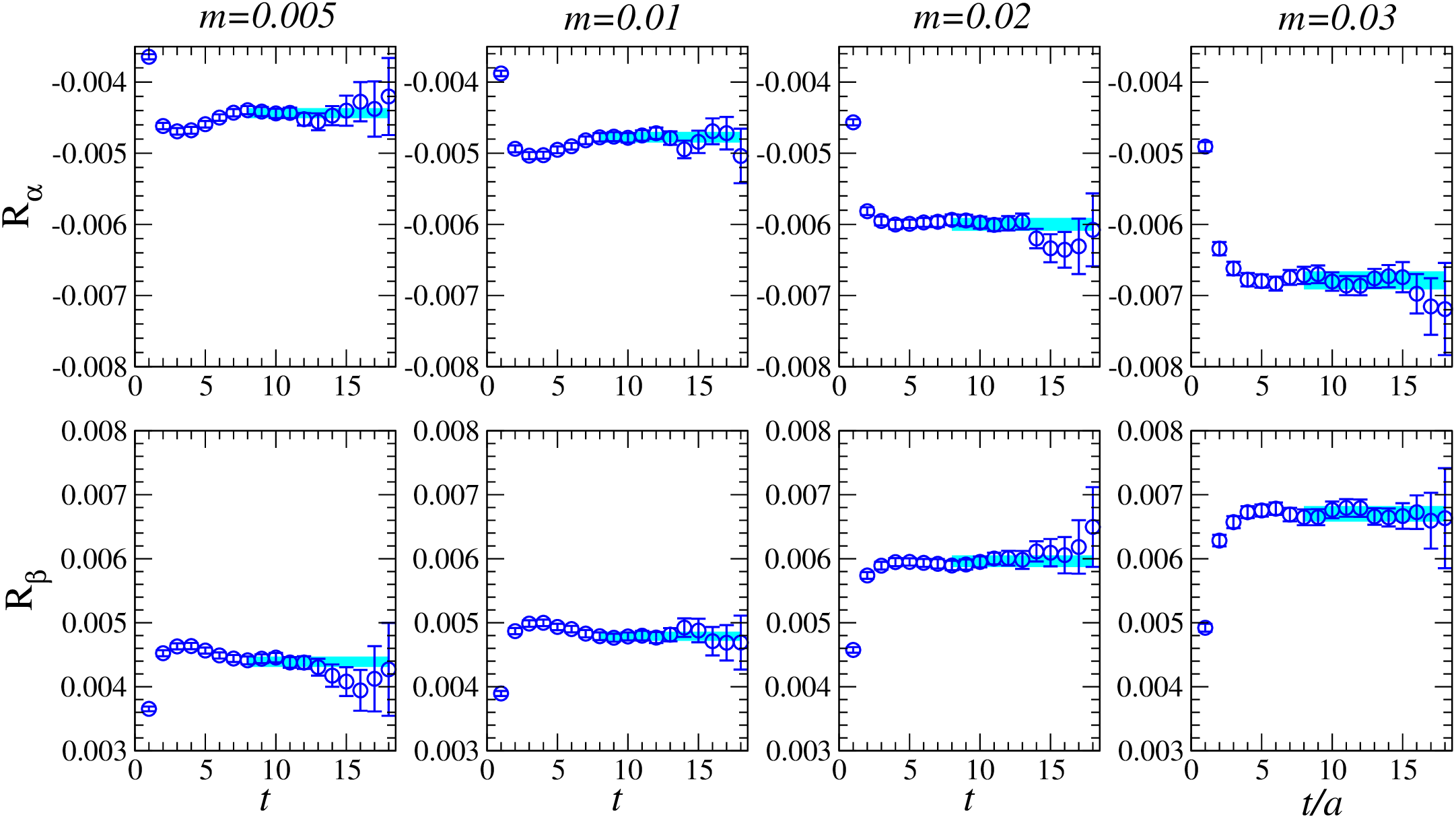}
\vskip 3mm
\caption{$R_\alpha(t)$ (upper) and $R_\beta(t)$ (lower) as a function of lattice time-slice.
  The cyan band denotes the line and statistical error of constant fitting
  within fitting range.
}
\label{fig:r_ave}
\end{center}
\end{figure}

%Figure \ref{fig:ralpha} shows that $R_\alpha(t)$ has the plateau region starting
%from $t/a=8$ at each quark mass.
Figure \ref{fig:r_ave} plots our result of $R_\alpha(t)$ and $R_\beta(t)$ obtained after averaging those with two different nucleon interpolating operators $\mathcal N_4$ and $\mathcal N_{45}$. Fitting to plateau is done to the range $t\in [8,18]$ for all cases as we have shown the straight bar in Figure \ref{fig:r_ave}, where the statistical error is included. 

\begin{figure}
\begin{center}
  \includegraphics[width=70mm]{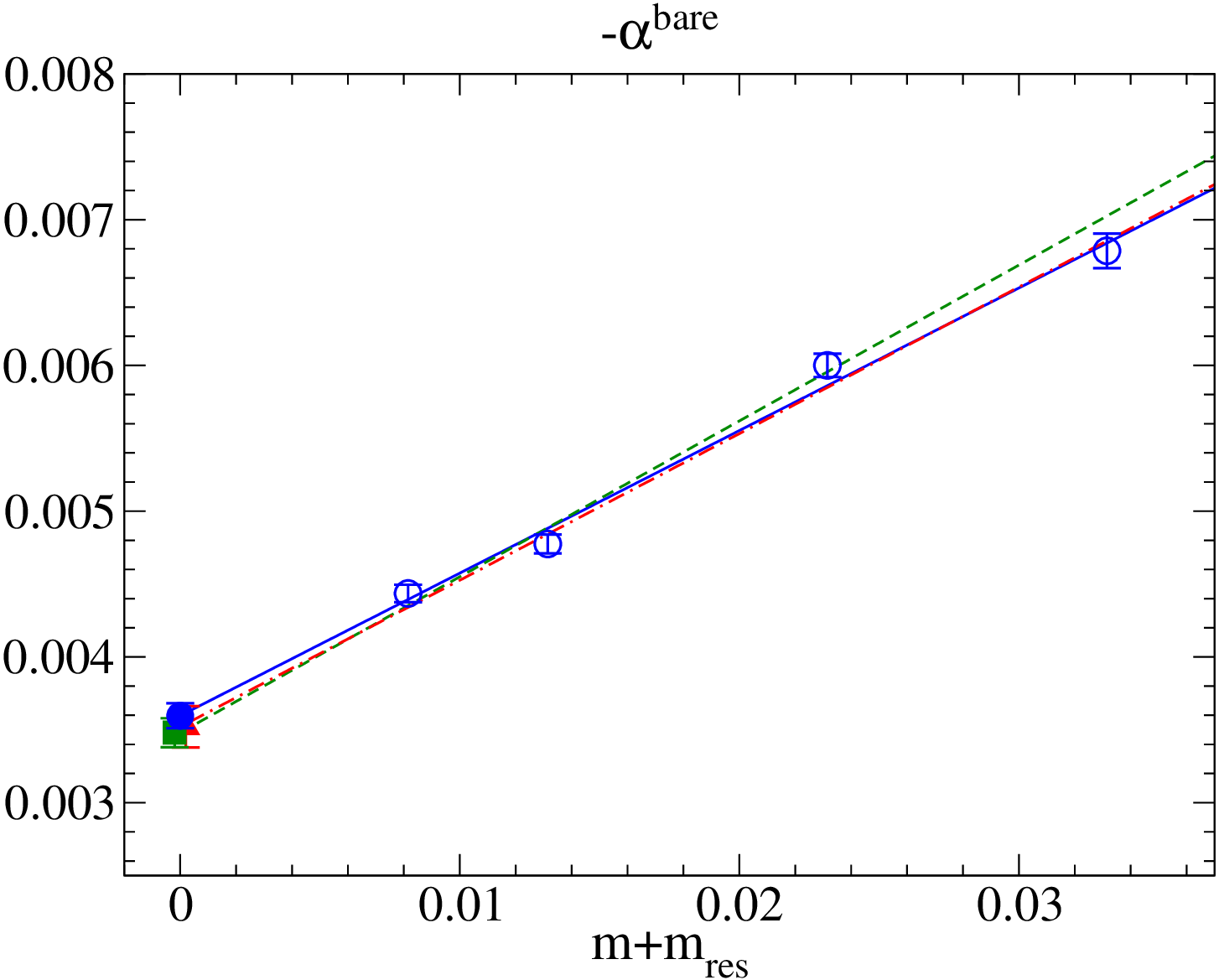}
  \hspace{3mm}
  \includegraphics[width=70mm]{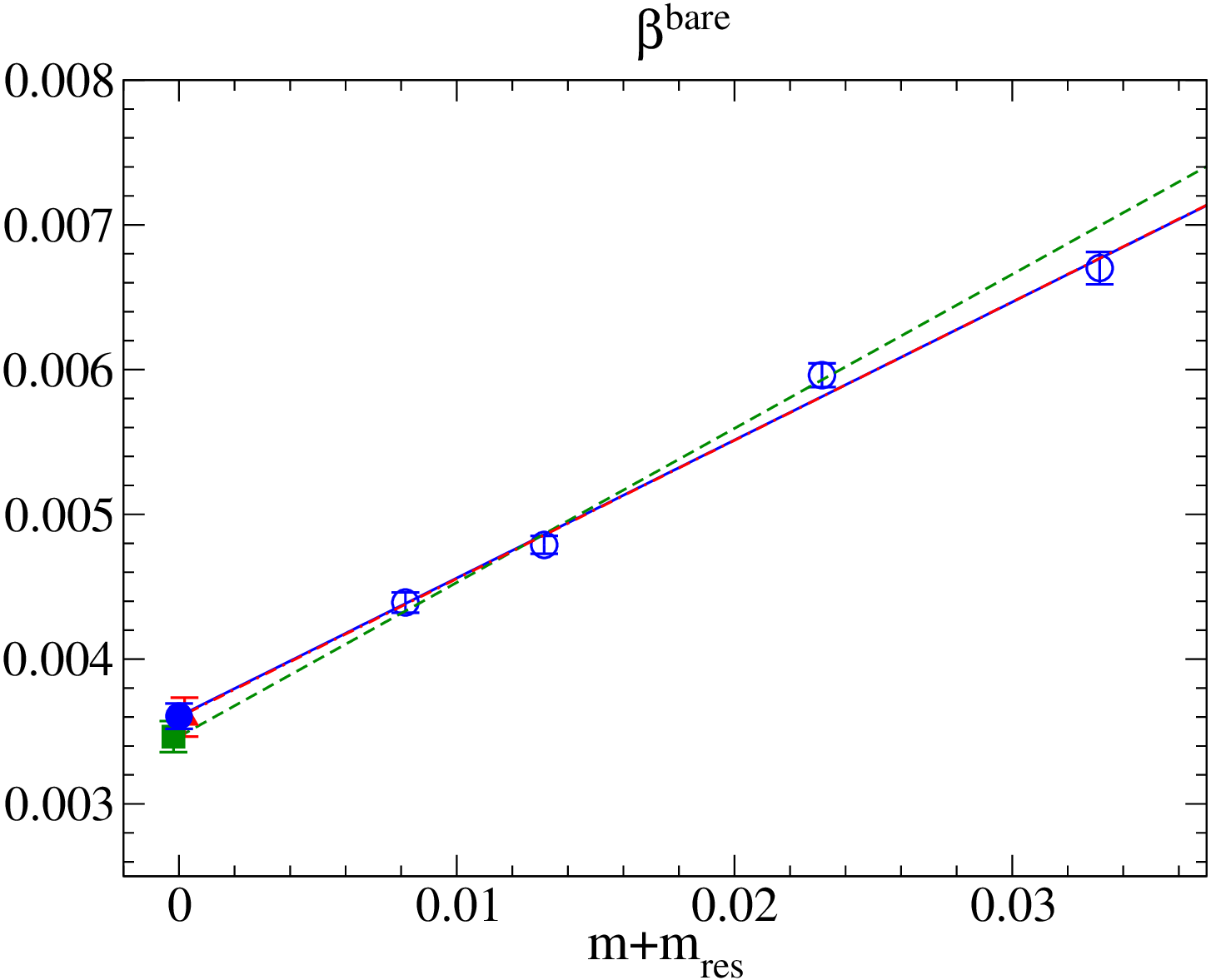}
  \vskip 3mm
\caption{Quark mass dependence of bare $\alpha$ and $\beta$. The
 straight lines are fits to the linear ansatz with three different
 fitting ranges: (i) containing all data [solid line (blue)], (ii)
 excluding the heaviest point [dashed-dotted line (red)] and (iii)
 excluding the lightest point [dashed line (green)] (see the text for
 detail). Filled symbols represent the values in the chiral limit from
 the three fits.  
}
\label{fig:r_mdep}
\end{center}
\end{figure}

Figure \ref{fig:r_mdep} shows the quark mass dependence of bare value of $\alpha$ and $\beta$. We observe that lattice data behaves as a linear function in our quark mass region, and the chiral extrapolation to physical quark mass is carried out with linear function of quark mass,
\begin{equation}
  f(\tilde m) = c_0 + c_1 \tilde m,
  \label{eq:lin_lecs}
\end{equation}
with $\tilde m = m + m_{\rm res}$, where the residual mass has been estimated as $m_{\rm res}=0.003152(43)$ \cite{Aoki:2010dy}. The (bare) physical quark mass has been obtained as
\begin{equation}
  \tilde m_{ud}^{\rm phys} = 0.001382, \label{eq:m_phys}
\end{equation}
from renormalized one $m^{\rm phys}_{ud}=-0.001770(79)$ \cite{Blum:2014tka}.

To estimate the uncertainty due to using the linear extrapolation, we use three different fitting ranges: (i) $m\in[0.005,0.03]$ (3.4,2.7), (ii) $m\in[0.005,0.02]$ (2.5,1.1) and (iii) $m\in[0.01,0.03]$ (2.5,2.5), where the number in the brackets show the resultant $\chi^2$/dof for $\alpha$ and $\beta$ respectively. The systematic error due to the assumption of linear behavior is evaluated from the maximum difference of central value between (i) and (ii), and (i) and (iii). The results are tabulated in Table~\ref{tab:lecs}. Compared to the previous work~\cite{Aoki:2008ku}, the statistical error has been improved to 2\% from 10\%, and systematic error of the chiral extrapolation is improved to 3\% from 20\%. Although the error estimation procedure is same as the previous work, because of the highly statistical precision we can use, the systematic error is properly estimated. 
%In other words the previous work overestimated the systematic error of the chiral extrapolation.

We estimate the systematic error of lattice artifact as 5\%, which is evaluated from comparison with different lattice spacing for hadron spectrum (see \cite{Aoki:2010dy}). The uncertainty in renormalization factor, which is dominated by the truncation error of the perturbative matching and running beyond the next-to-leading order. This error turns out to be the most dominant error.

The final value at $\mu=2$ GeV in $\overline{\rm MS}$ NDR scheme \ref{eq:zfact} extrapolated to physical quark Eq.~(\ref{eq:m_phys}) is 
\begin{equation}
  \alpha = -0.0144(3)(21)\,{\rm GeV^3}, \quad \beta = 0.0144(3)(21)\,{\rm GeV^3}, 
  \label{eq:lecs}
\end{equation}
where the first error is statistical and the second is systematic
obtained by the quadrature. The total error is around 15\%, which is
improved from 22\% \cite{Aoki:2008ku}.  
The superficial relation $\alpha+\beta=0$ is observed. 
The relation should hold for the non-relativistic limit and the
approximate relation is known to hold at least numerically in the
quenched case \cite{Aoki:2006ib}.  Here we have confirmed that it holds in the $N_f=2+1$
case with an improved precision.
Using these low energy constants, the relevant form factor can be
computed via BChPT formula (see Appendix~\ref{sec:bchpt}). 

\begin{table}
\begin{center}
\caption{Error budget of $\alpha$ and $\beta$ at $\mu=2$ GeV. ``$\chi$'' column is systematic uncertainty due to chiral extrapolation.''$a^2$'' column denotes the systematic uncertainty due to $\mathcal O(a^2)$. ``$\Delta_Z$'' and ``$\Delta_a$'' column are systematic uncertainties from renormalization factor and lattice spacing.}
\label{tab:lecs}
\begin{tabular}{c|c|cccc}
\hline\hline
\multirow{2}{*}{LECs} & \multirow{2}{*}{statistical error} & systematic error \\\cline{3-6}
 & & $\chi$ & $a^2$ & $\Delta_Z$ & $\Delta_a$ \\
\hline
$\alpha$(GeV$^3$) = $-$0.0144(15) & 0.0003 & 0.0005 & 0.0007 & 0.0012 & 0.0002 \\
$\beta$(GeV$^3$) = 0.0144(15) & 0.0004 & 0.0005 & 0.0007 & 0.0012 & 0.0002 \\
\hline\hline
\end{tabular}
\end{center}
\end{table}

\section{Improved result of relevant form factor}\label{sec:ff}
In this section, we show our improved result of the form factors from ``direct'' measurement in which we compute the three-point function of $N\rightarrow P$ including baryon number violating operator. Compared to our previous study \cite{Aoki:2013yxa}, the results are improved by the use of the AMA technique. We also add one larger meson momentum point, $n_p=(1,1,1)$ to the two non-zero momentum we had, $n_p=(1,0,0)$ and $n_p=(1,1,0)$. By that we now are able to estimate the systematic error from $\mathcal O(q^4)$ term.

\begin{figure}
\begin{center}
\includegraphics[width=100mm]{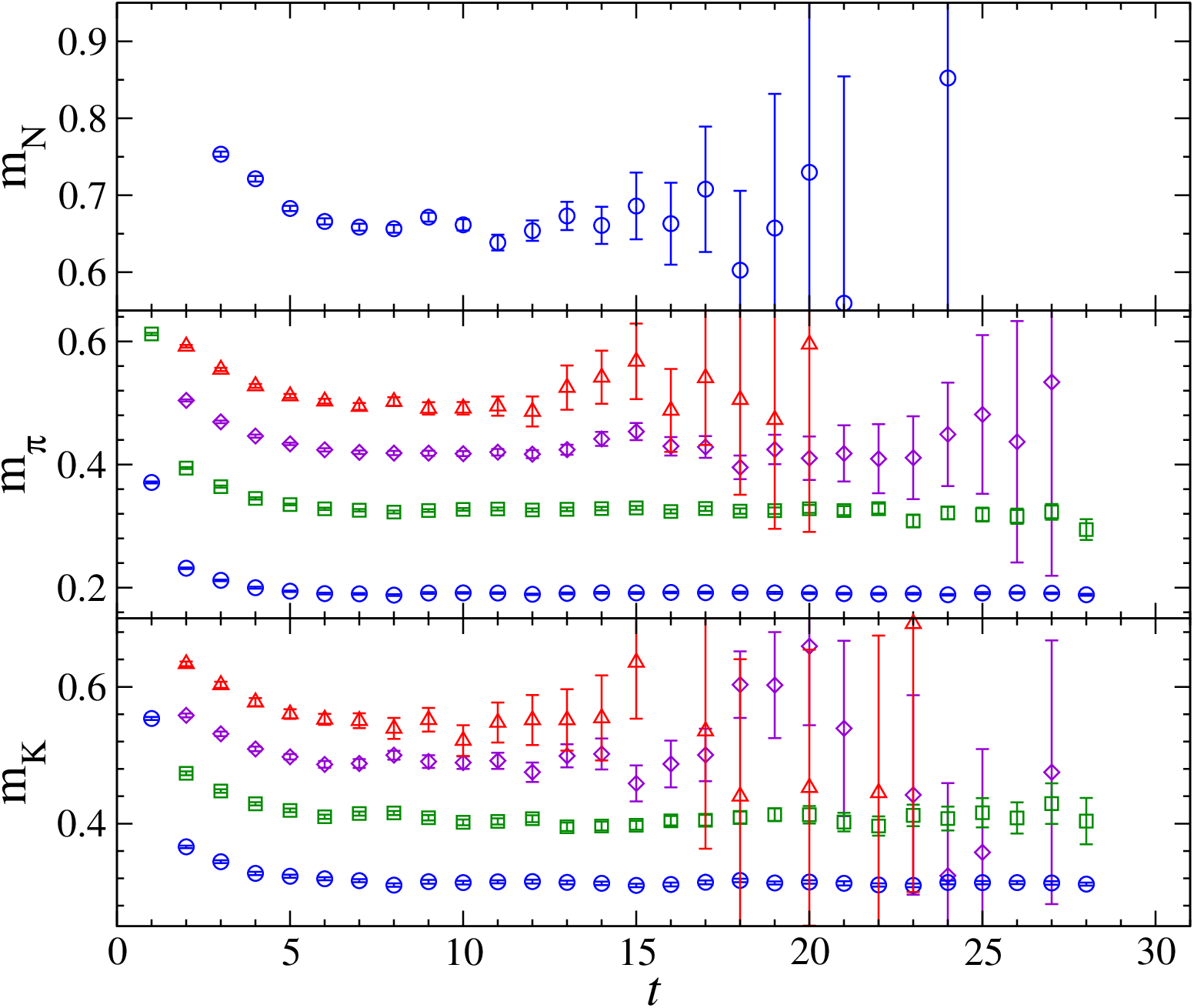}
\caption{Effective mass of nucleon, pion and kaon from the top to bottom panel.
  Different symbols denote the data with variation of momentum as
  zero (circle), $\vec n_p=(1,0,0)$ (square), $\vec n_p=(1,1,0)$ (diamond) and $\vec n_p=(1,1,1)$ (triangle).
  The quark mass is $m=0.005$.}
\label{fig:effm}
\end{center}
\end{figure}

In Figure \ref{fig:effm}, we plot the effective mass of nucleon, pion and kaon with momentum we use in the construction of ratio Eq.~(\ref{eq:ratio}). One clearly sees the plateau starting from $t=6$ in those hadrons, so from here we regard that the ground state is dominant from $t=6$. 

\begin{figure}
\begin{center}
\includegraphics[width=110mm]{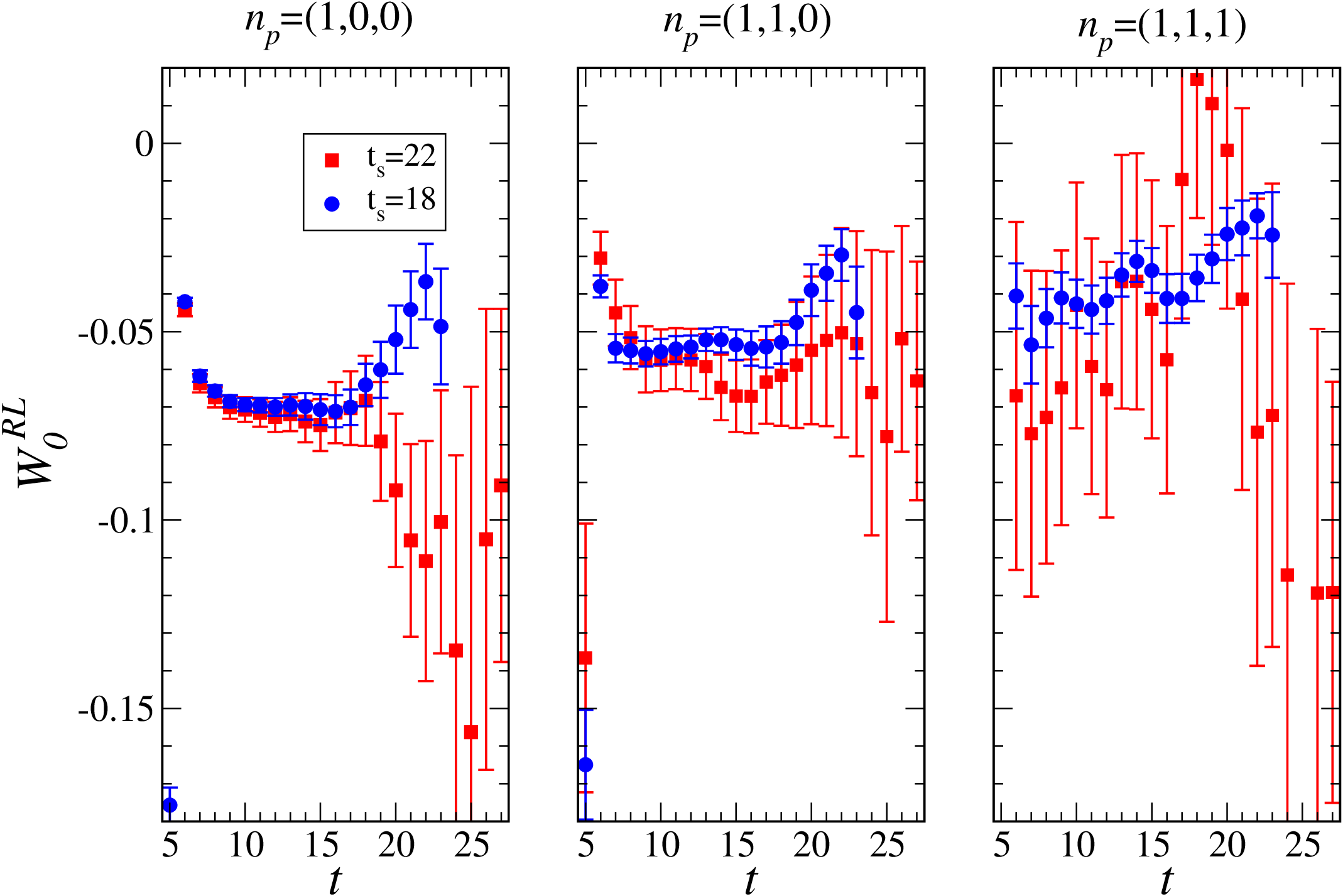}
\vskip 5mm
%\hspace{5mm}
\includegraphics[width=110mm]{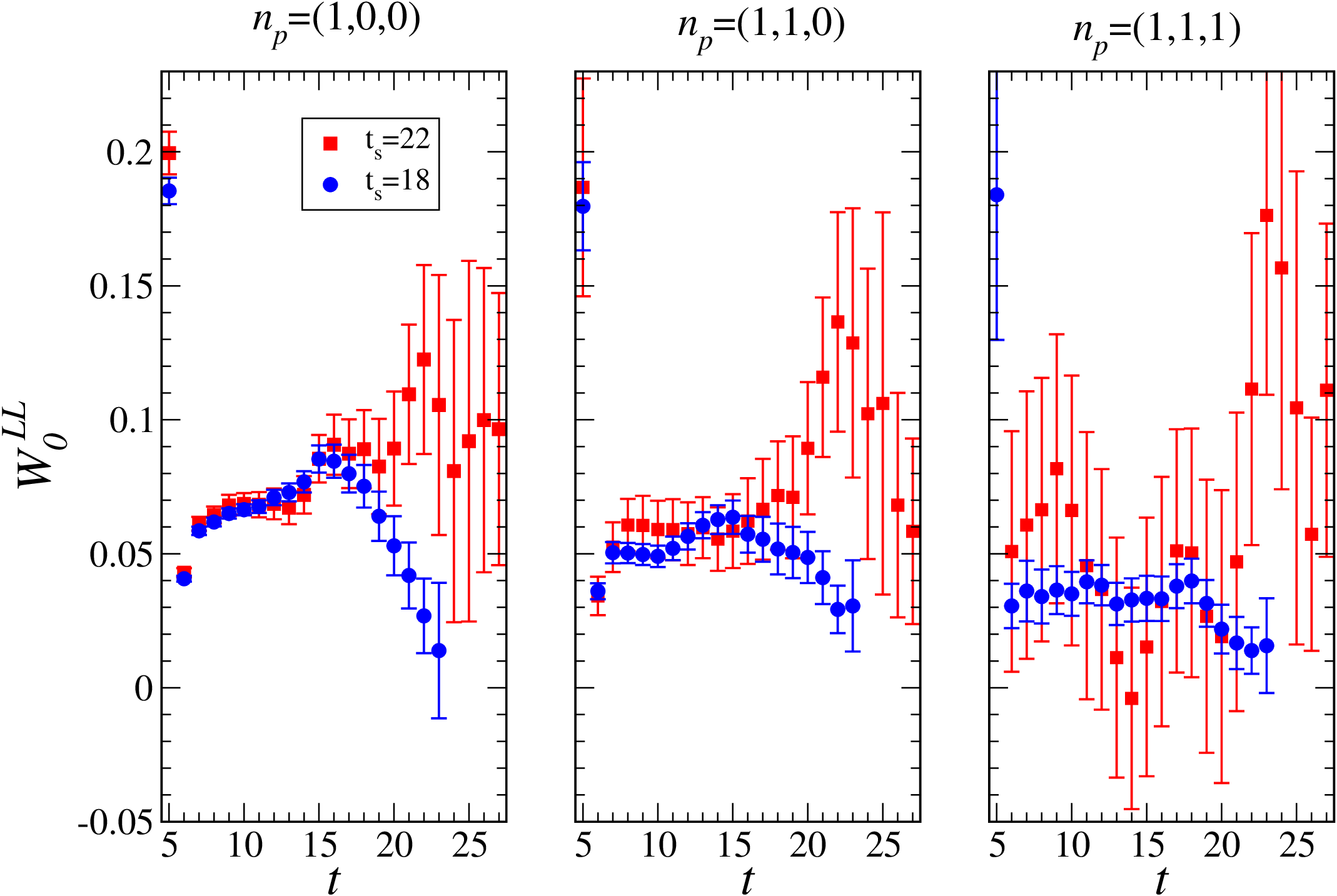}
\vskip 3mm
\caption{
Bare $W_0(t)$ for $p\rightarrow \pi^0$ transition with $\Gamma=R$ (top) and $\Gamma=R$ (bottom) in $m=0.005$. Different symbols are results with long $t_s=22$ (red) and short $t_s=18$ (blue). Left, middle and right panels are result at momentum $\vec n_p=$(1,0,0), (1,1,0) and (1,1,1) respectively. Source nucleon ($t = 5$)and sink pion location (27 or 23) with separations 22 and 18 respectively.
%  separated as $t_s=22$ (18) is 5 and 27 (5 and 23) respectively.
%\red{POSITIVE SIGN OF CAPTION X-AXIS}
}
\label{fig:w0_tsdep}
\end{center}
\end{figure}

Figure \ref{fig:w0_tsdep}, in which we plot the time-slice dependence of the matrix element extracted from the ratio Eq.~(\ref{eq:ratio}) for the $p\to\pi^0$ mode at our lightest point $m=0.005$, shows the comparison with two different source-sink separations, $t_s=18$ and 22 corresponding to $t_s=1.98$ and 2.43 fm respectively. The time separation $t_s=18$ is new in this study with four-time slice shorter than original $t_s=22$ \cite{Aoki:2013yxa}. We observe plateau for shorter separation at $t\in[12,16]$, where the denominators are also dominated with the ground state (see Fig.~\ref{fig:effm} and note that the source is located at $t=5$ here). The plateaus from two separations are consistent and the shorter separation yields significantly better statistical accuracy. Let us note that the clear plateau is observed even in largest momentum case $n_p=(1,1,1)$.
%This is made possible only with the AMA method.
Hereafter we only use the result in $t_s=18$, and further test the effect of excited state contamination by changing the fitting range below.

\begin{figure}
\begin{center}
  \includegraphics[width=70mm]{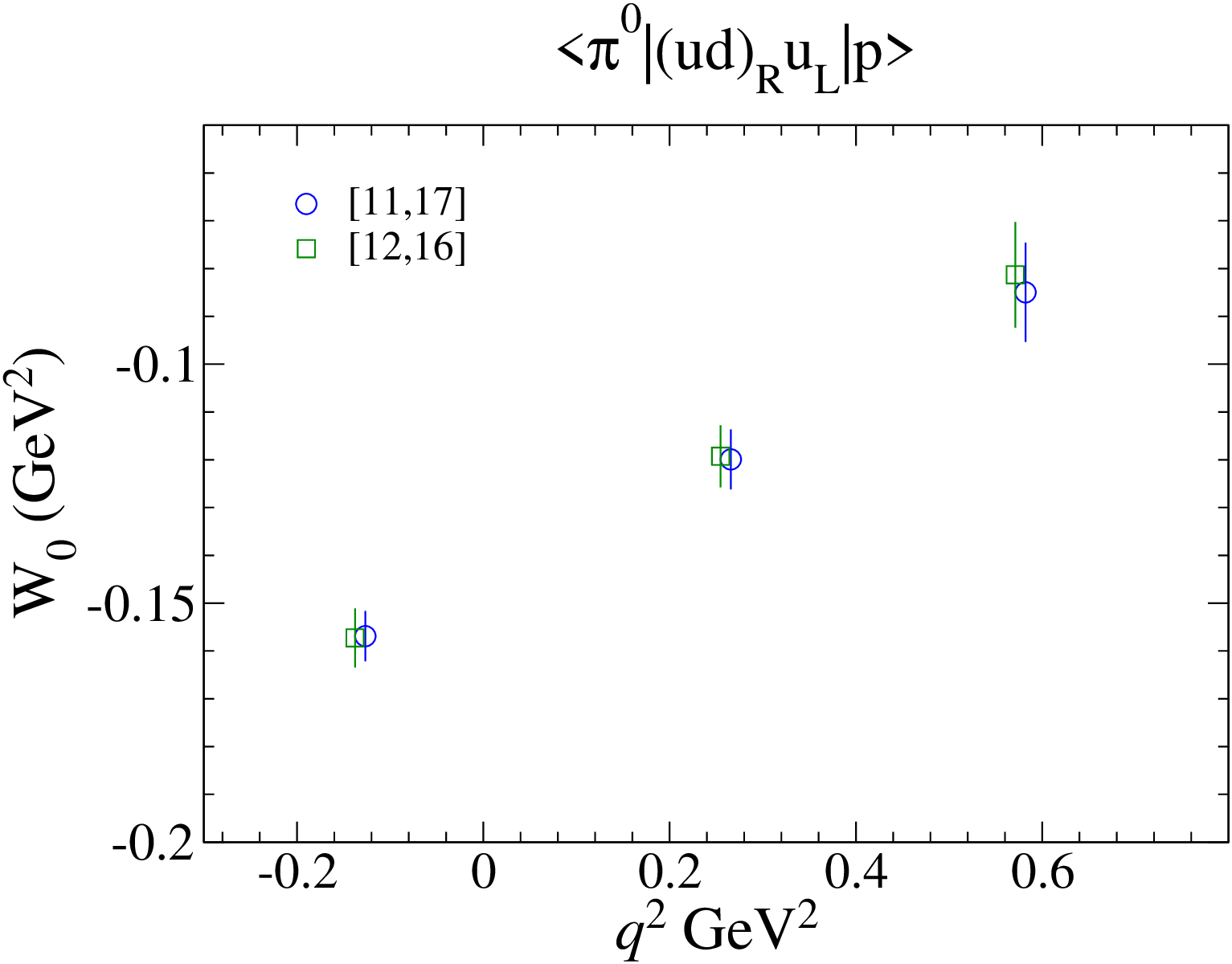}
  \hspace{3mm}
  \includegraphics[width=70mm]{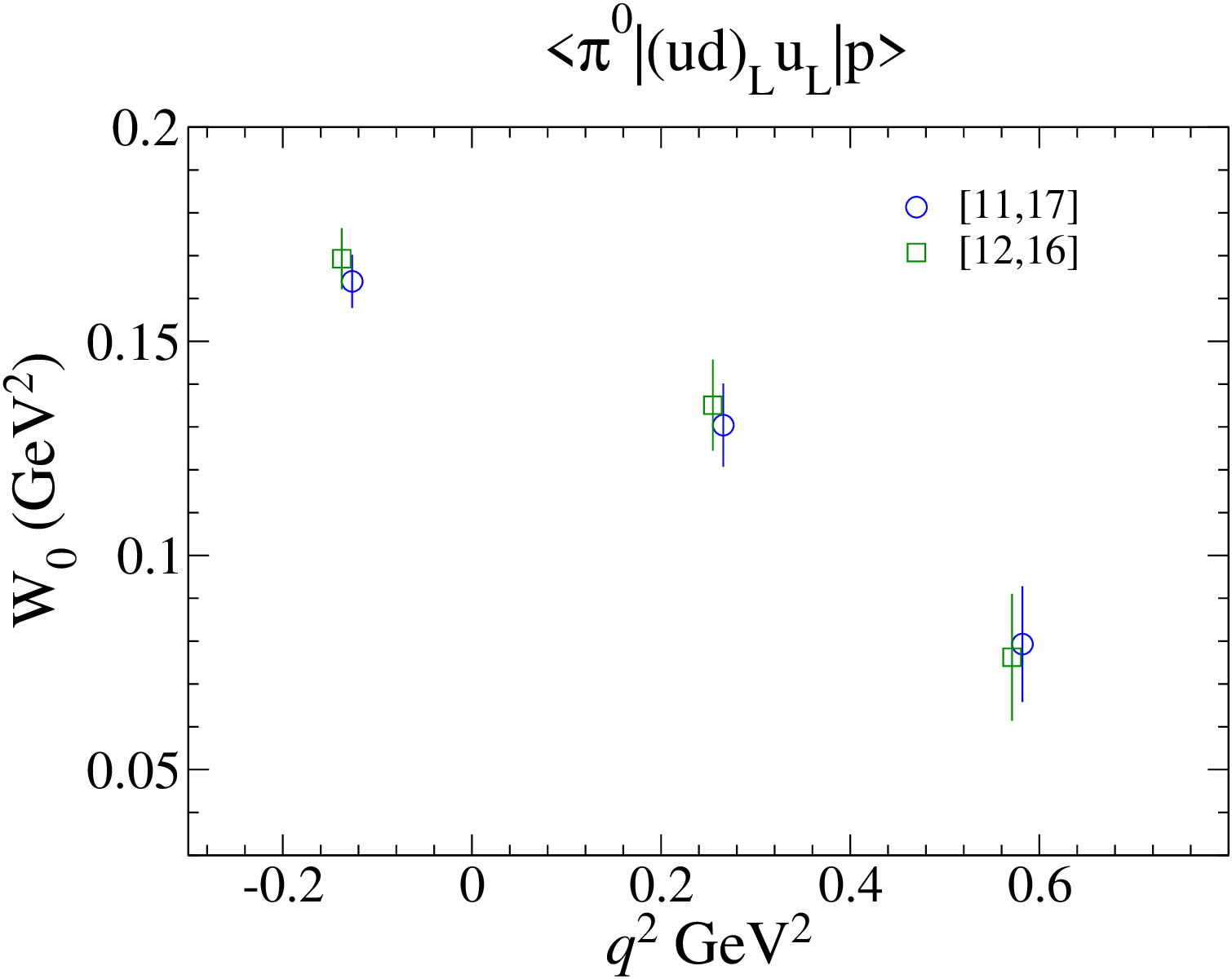}
  \vskip 3mm
  \caption{Renormalized $W_0$ obtained by plateau fit for data in Figure \ref{fig:w0_tsdep} with three fitting ranges, $[13,17]$, $[11,17]$ and $[10,18]$ at $m=0.005$. Left panel is a result for $\Gamma=R$ and right is a result for $\Gamma=L$ in $p\rightarrow\pi$ channel.}
\label{fig:w0_fitdep}
\end{center}
\end{figure}

Figure \ref{fig:w0_fitdep} shows the result of plateau fit for $W_0$ using variation of fitting ranges as $t\in[11,17]$ and $t\in[12,16]$ to study the effect of excited state contamination into the signal. We observe those values are consistent within 1 sigma error in each $q^2$, while the central value has slight tension, especially for the lowest momentum in $\Gamma=L$. In order for estimate of systematic uncertainties including the effect of excited state contamination, we compare the results using those fitting ranges. We will back to such a discussion later. 

\subsection{Global fitting}\label{sec:global}
To perform the extrapolation to kinematic point and physical pion mass simultaneously, we globally fit all lattice data with the linear ansatz for quark mass and $q^2$ dependence as
\begin{equation}
  F_{W_0} = A_0 + A_1\tilde m + A_2 q^2, 
  \label{eq:fw0}
\end{equation}
where $\tilde m$ is the same definition as in eq.(\ref{eq:lin_lecs}). Figures \ref{fig:w0_q2dep_R} and \ref{fig:w0_q2dep_L} plot renormalized $W_0(q^2)$ for every decay channel in each quark mass. We observe that lattice data in each quark mass, which denotes the same symbols in Figures \ref{fig:w0_q2dep_R} and \ref{fig:w0_q2dep_L}, is behaved as linear $q^2$ dependence. For mass dependence, we also observe the monotonic decreasing or increasing when $m$ is increasing. Even using the linear ansatz $\chi^2$/dof is reasonably small (note that we use uncorrelated fits) as presented in the first ``$\chi^2_{\rm dof}$'' column of Table \ref{tab:w0_chiral}.

We next study the uncertainties in the fitting related with the mass dependence, following the method used in Ref.~\cite{Aoki:2013yxa}. The estimated errors are attributed to the higher order correction than $\mathcal O(m)$ and a part (at least) of the finite volume effect. Table~\ref{tab:w0_chiral} presents the errors estimated with the discrepancy from central value, which is obtained by full range, $m\in[0.005,0.03]$, and two fitting ranges; $m\in[0.005,0.02]$ for ``light'' region and $m\in[0.01,0.03]$ for ``heavy'' region. The error with``light'' region can be an estimate of $\mathcal O(m^2)$ correction since exclusion of heavy mass makes less $\mathcal O(m^2)$ correction. On the other hand, the error with ``heavy'' region can (at least partly) be due to the finite volume effect, since the lightest point suffers most from the effect with the fixed volume. In each range, $\chi^2$/dof is not significantly large. The values presented in the table are taken as the maximum error compared with the result obtained in two $t$ fitting ranges $t\in[12,16]$ and $t\in[11,17]$. %\footnote{In practice we take the maximum separation in four values obtained by two ranges in $t$ and two ranges in $m$.}.
% For $p\rightarrow\pi$ channel, one
% sees that $RL$ chirality is largely affected from $\mathcal O(m^2)$
% rather than $\mathcal O(e^{-m_\pi L})$ terms, while $LL$ has similar
% effect. In addition, the magnitude of such a discrepancy between $LL$
% and $RL$ chirality for $p\rightarrow\pi$, $LL$ has larger discrepancy
% than $RL$. In $p\rightarrow K$ channel, we also observe similar
% tendency.
The ``total'' error of the chiral extrapolation in the table is calculated by adding two errors, ``light'' and ``heavy'', in quadrature.
%({\it \color{red} YA: if the ``total'' error is calculated through
%quadrature, it seems obvious double counting happening for $t$-range
%error???})

In similar manner as ``light'' error, the $\mathcal O(q^4)$
error is estimated from the difference of the results obtained with the
full range of $q^2$ with all the non-zero meson momentum and the shorter
range where largest $|q^2|$ ($n_p=(1,1,1)$) is neglected. The result is
shown in the column labeled as $\mathcal O(q^4)$ in Table~\ref{tab:w0_chiral}.
This error turns out to be smaller than that of the ``chiral'' extrapolation.

As shown in Figure~\ref{fig:w0_q2dep_R} and \ref{fig:w0_q2dep_L}, the $q^2$ dependence obtained by extrapolation of data in``direct'' method does not largely differ from BChPT including $\alpha$ and $\beta$ obtained in section \ref{sec:lecs}, especially for that $p\rightarrow \pi$ channel has a tendency to be close to each other when increasing $q^2 > 0$. There is a discrepancy up to about a factor of 2 around the kinematics point. Such a comparison will be discussed later.
%({\it \color{red} YA could not understand what this paragraph wants to describe. And we really discuss more details later ?})

\subsection{Sequential fitting}
\begin{figure}
\begin{center}
\includegraphics[width=120mm]{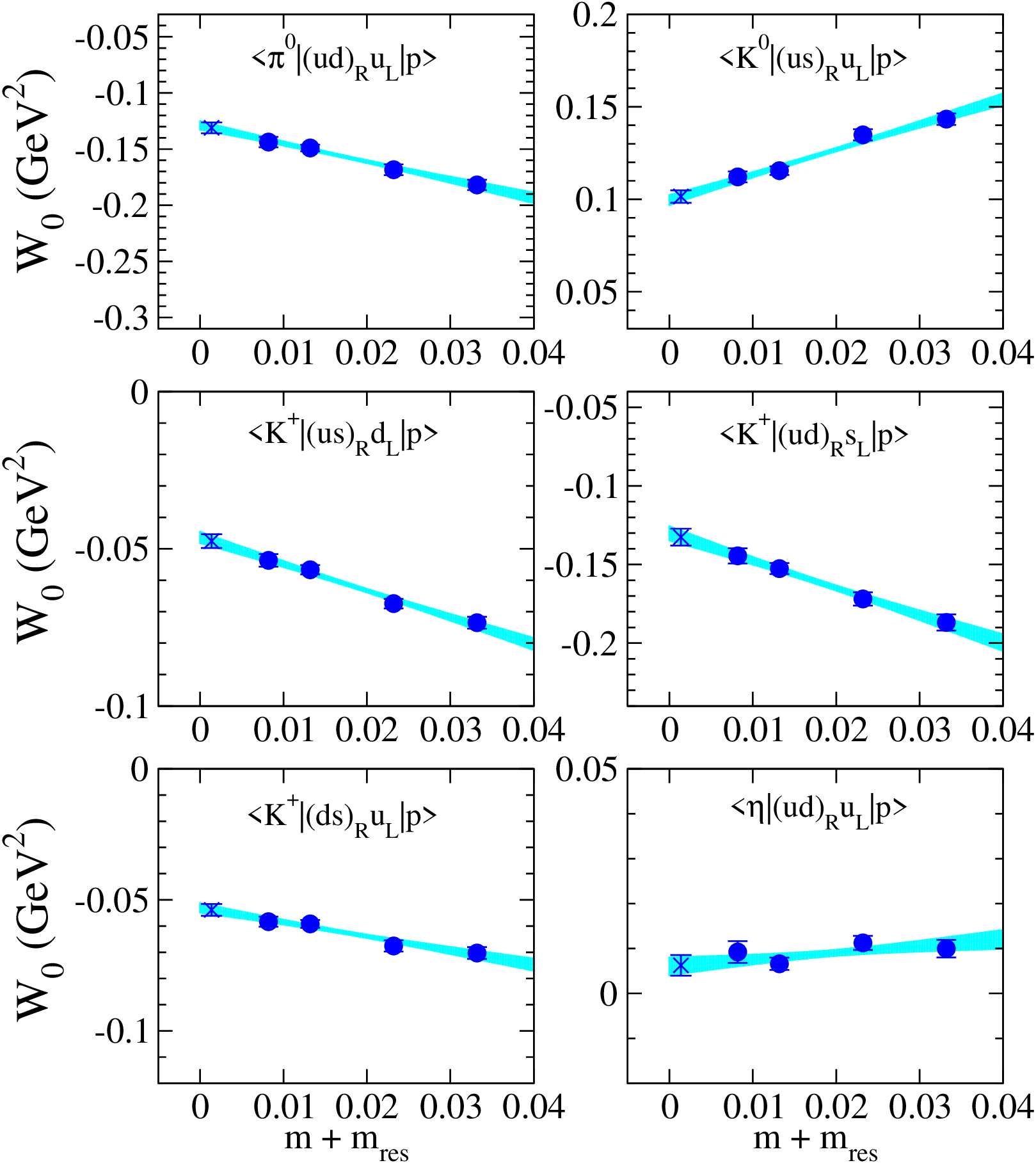}
\vskip 4mm
\caption{
Quark mass dependence of renormalized $W_0$ after $q^2=0$ extrapolation for $\Gamma=R$ at each channel. Cross symbol denotes $W_0$ in physical quark mass after linear extrapolation. Band is extrapolation line including statistical error.}
\label{fig:w0_mdep_R}
\end{center}
\end{figure}

\begin{figure}
\begin{center}
\includegraphics[width=120mm]{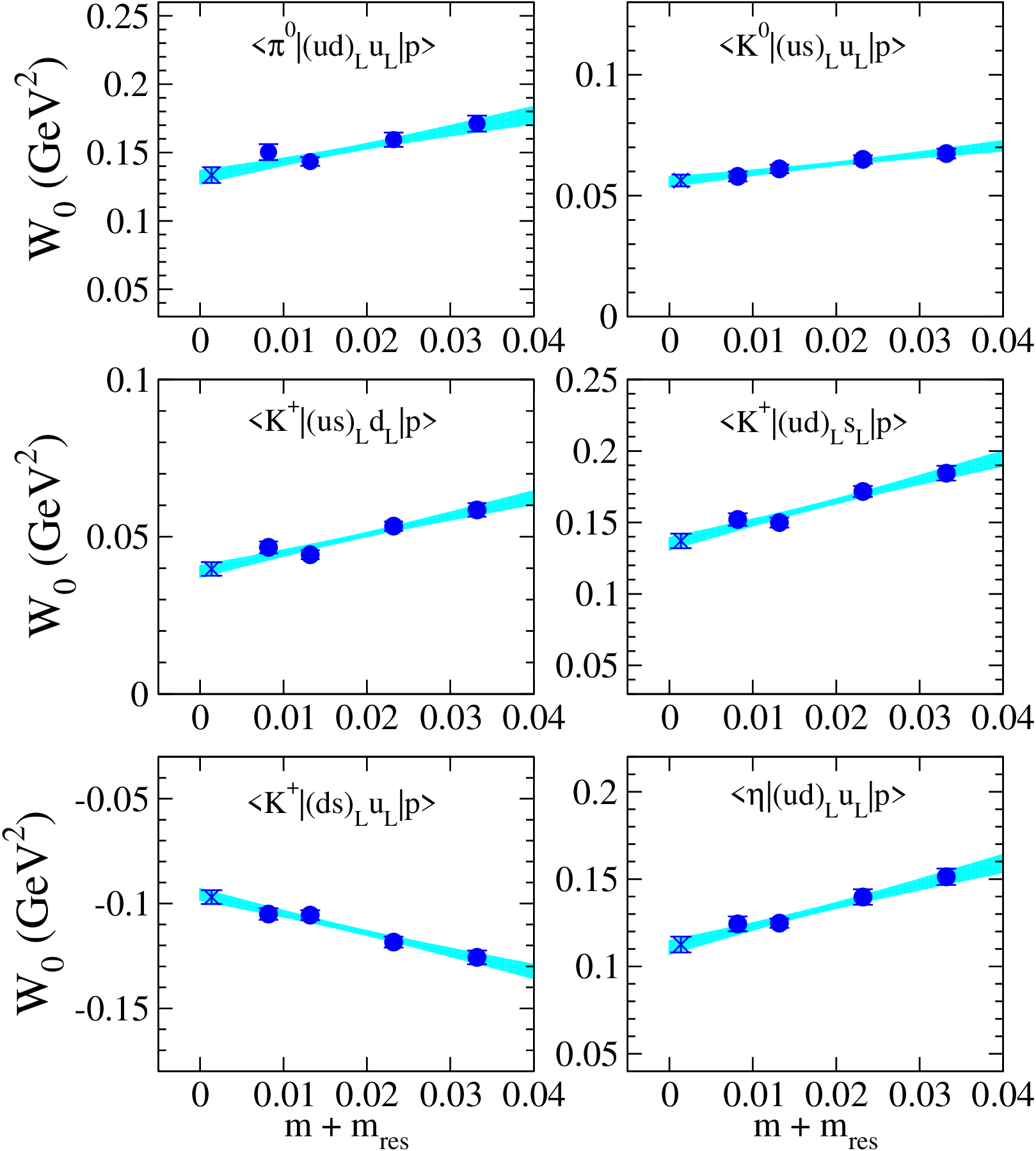}
\vskip 4mm
\caption{
Quark mass dependence of renormalized $W_0$ after $q^2=0$ extrapolation for $\Gamma=L$. Each symbol is same as Figure \ref{fig:w0_mdep_R}.
}
\label{fig:w0_mdep_L}
\end{center}
\end{figure}

In the global fitting we estimated a part of the systematic errors due
to omitting the higher order terms in the expansion of the light quark mass
$\tilde m$ and squared momentum transfer $q^2$. Those estimated are of $\mathcal
O(\tilde m^2)$ and $\mathcal O(q^4)$. 
Remaining error is of $\mathcal O(\tilde mq^2)$.  For the estimate we use the
same method as in Ref.~\cite{Aoki:2013yxa}.
The procedure is that first the $q^2\to 0$ extrapolation is carried out for
each fixed quark mass with linear function, then chiral extrapolation
is performed (see Figure \ref{fig:w0_mdep_R} and \ref{fig:w0_mdep_L}).
By doing that we are taking into account the $q^2$
dependence in prefactor of the linear quark mass term, $A_1$ in
Eq.~(\ref{eq:fw0}). 
If the result is different, it is attributed as the $m$ effect in $A_1$,
thus is of $\mathcal O(\tilde mq^2)$.
% We notice that data for $LL$
% chirality plotted in Figure \ref{fig:w0_mdep_L} shows the chiral
% behavior is slightly separated from linear function, especially for the
% lightest point. The $\chi^2$/dof (see the last column in
% Table~\ref{tab:w0_chiral}), which is from fitting with full mass range,
% becomes large in $LL$ chirality. In the previous section, comparison
% with ``heavy'' fitting range of global fit has also indicated such a
% tendency for $LL$ chirality.  
% %%%
% The last two columns of Table \ref{tab:w0_chiral} presents the $\chi^2$/dof using full mass range and relative error obtained by discrepancy from global fit result. $\chi^2$/dof value is still reasonable, and estimated $\mathcal O(mq^2)$ uncertainty is relatively small rather than chiral extrapolation error. 
The last column of Table \ref{tab:w0_chiral} shows the $\chi^2$ per degree of freedom for the final $\tilde m$ linear fit. The second last column represents the systematic error estimated in this analysis. It turns out to be sub-dominant in the fitting errors.

\subsection{The final results}
Table \ref{tab:w0_error} presents the summary of the nucleon decay form
factor for each operator and final state with the statistical and
systematic errors. The statistical error is significantly reduced to
1/4--1/6 from our previous study~\cite{Aoki:2013yxa} and now is sub-dominant.
The systematic errors for the extrapolation discussed above are combined and shown in the ``$(\tilde mq^2)$-fit'' column. Since we use a single lattice cutoff in this study, the lattice artifact, which is $\mathcal O(a^2)$ correction, is estimated from the scaling study of hadron spectrum as done in \cite{Aoki:2010dy}. The mass of valence strange quark which participate in the matrix elements of kaon final state is set equal to its sea-quark mass $m=0.04$. There is a mismatch to the physical strange mass. The associated systematic error is estimated using a subset of the $m=0.005$ ensemble by setting $m=0.343$ \footnote{the value comes from the physical strange quark mass used in the previous study. The latest estimate of physical strange mass \cite{Blum:2014tka} turns out to be 0.03224 which is not far enough to change the systematic error estimate.}. The difference of central value is by 3 \% at most.  We conservatively take 3 \% as the systematic error of the form factors for the process with the kaon final state due to the use of the mismatched strange sea and valence quark masses. On the other hand, the mismatch effect of the strange sea quark is expected to be much less than that of the valence quark, thus, it is negligible in pion and eta final state. The largest uncertainty comes from that of the renormalization factor, which is dominated by the systematic error due to the truncation of the perturbative matching (Eq.~(\ref{eq:zfact}). The total error summing up all in quadrature amounts to 10--15 \% for the form factors with pion and kaon final state.

Additionally, Table \ref{tab:wmu} presents the matrix element with muon final state, $m_l=m_\mu$. $W_\mu$ in Eq.(\ref{eq:w_mu}) is made from two form factors, $W_0$ and $W_1$. As one sees in Figure \ref{fig:w1_q2dep_R} and \ref{fig:w1_q2dep_L}, the magnitude of $W_1$ in each matrix element is similar to $W_0$, and hence $W_1$ term multiplied with factor $m_\mu/m_N\sim 0.1$ in $W_\mu$ affects around 10\% effect to matrix element in the kinematics with muon final state. 

Note that for matrix element with eta final state we are ignoring the
disconnected diagram, which means there remains additional
uncertainty. However, the contribution of disconnected diagram expects
to be small from OZI suppression. Detailed study in eta sector including
disconnected diagram is beyond the scope of this paper. 

\begin{figure}
\begin{center}
  \includegraphics[width=120mm]{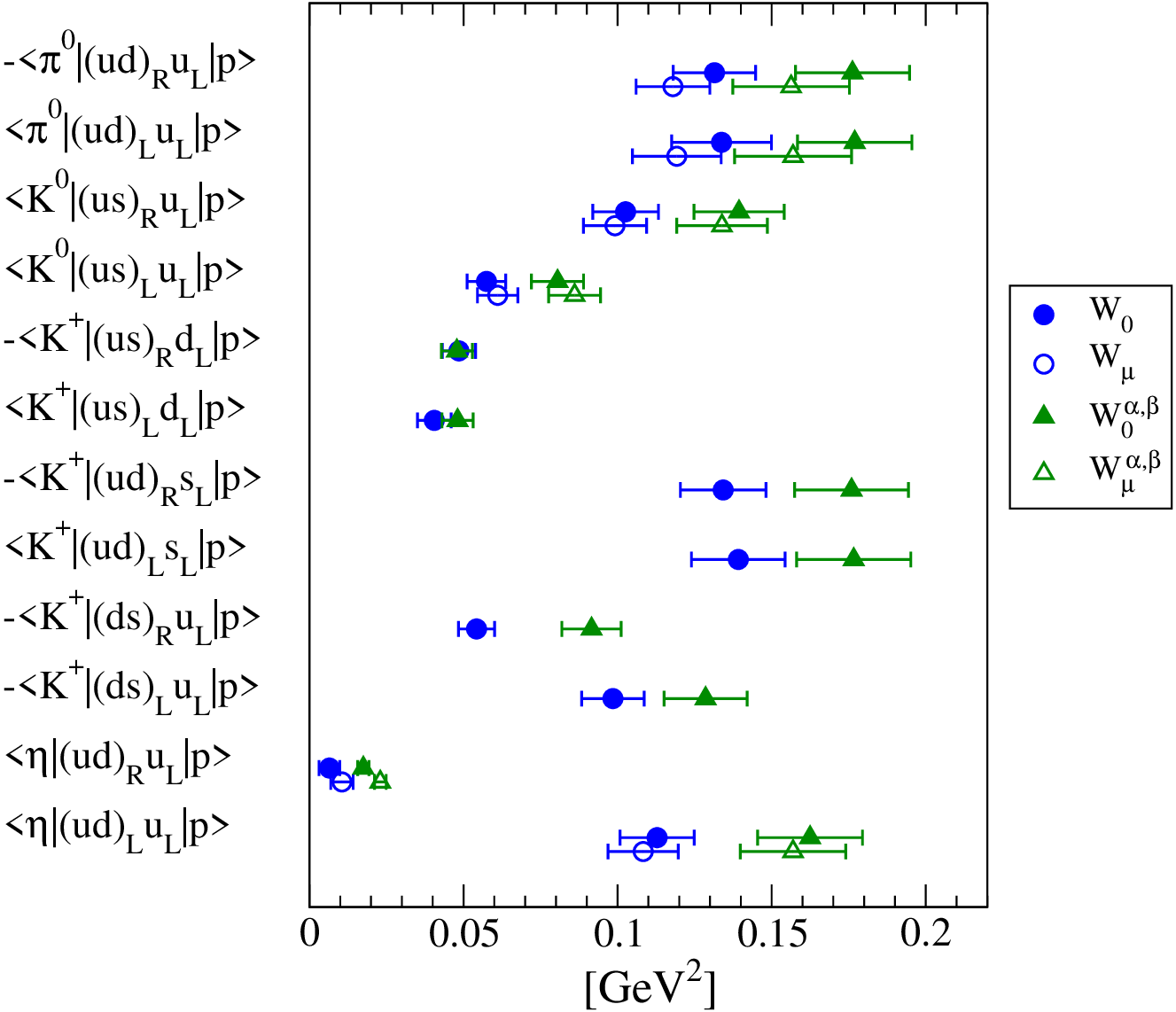}
  \vskip 2mm
\caption{Summary of matrix elements obtained in our study; ``$W_0$, $W_\mu$'' which are evaluated from ``direct'' method and ``$W^{\alpha,\beta}_0$, $W^{\alpha,\beta}_\mu$'' which are evaluated ``indirect'' method, including the systematic error as discussed in text. 
}
\label{fig:w0_sum}
\end{center}
\end{figure}

All the final results of the relevant form factors of proton decay
$W_0$ and $W_\mu$ with the ``direct'' method are summarized in
Figure~\ref{fig:w0_sum}. 
The results are also compared with those with the ``indirect'' method
through BChPT using lattice LECs (denoted as $W^{\alpha,\beta}_0$ and
$W^{\alpha,\beta}_\mu$).
The ``indirect'' method always overestimates the form factor.
The amount is 25\% or more except for two cases ($\langle
K^+|(us)_{R/L}d_L|p\rangle$). 
In contrast to previous study~\cite{Aoki:2013yxa}, each error becomes lot
smaller, and now we clearly see the discrepancy between $W_{0,\mu}$ and
$W^{\alpha,\beta}_{0,\mu}$ for most cases.

The fact that the indirect method which uses BChPT works poorly
is understandable as the physical kinematical point for the outgoing
pion is far from the soft pion limit, where the ChPT description becomes
arbitrary precise. We tested the soft pion theorem for the form factors
of the pion final state, which is found in the appendix \ref{appx:soft}. 
There the results from the indirect and direct
method appears to be consistent with each other in the soft pion limit.

%\red{SYSTEMATIC ERROR IS REDUCED BY IMPROVED PRECISION AND INCREASING MOMENTUM.}

\begin{table}[t]
\begin{center}
\caption{
Relative error of systematic uncertainty in chiral extrapolation estimated from three fitting ranges; ``light'' is fitting range without the heaviest point, ``heavy'' is fitting range without the lightest point. ``total'' is total one in quadrature. For reference, we also show the value of chi-squared per degree-of-freedom in our fitting as in ``$\chi^2_{\rm dof}$'' column. 
%\red{ADD THE ANALYSIS OF O($Q^4$)}
}
\label{tab:w0_chiral}
\begin{tabular}{r|cc|cc|cc|cc|cc}
\hline\hline
\multirow{2}{*}{Matrix element} & \multicolumn{5}{c}{Relative error in chiral extrapolation} & & & & \\
& total & $\chi^2_{\rm dof}$ & light & $\chi_{\rm dof}^2$ & heavy & $\chi^2_{\rm dof}$
& $\mathcal O(q^4)$ & $\chi^2_{\rm dof}$ & $\mathcal O(mq^2)$ & $\chi^2_{\rm dof}$ \\
\hline
$\langle \pi^0|(ud)_Ru_L|p\rangle$ & 1.8\% & 0.6 & 1.6\% & 0.8 & 0.8\% & 0.8 & 0.7\% & 0.6 & 0.3\% & 0.2 \\
$\langle \pi^0|(ud)_Lu_L|p\rangle$ & 5.7\% & 1.4 & 3.8\% & 2.0 & 4.3\% & 1.2 & 2.3\% & 2.2 & 2.6\% & 1.9 \\
\hline
$\langle K^0|(us)_Ru_L|p\rangle$ & 2.8\% & 1.4 & 2.7\% & 1.7 & 0.7\% & 1.5 & 0.7\% & 1.6 & 1.1\% & 1.4 \\
$\langle K^0|(us)_Lu_L|p\rangle$ & 3.1\% & 1.7 & 0.8\% & 1.9 & 3.0\% & 1.7 & 1.0\% & 2.0 & 2.1\% & 0.2 \\
$\langle K^+|(us)_Rd_L|p\rangle$ & 3.5\% & 1.3 & 3.4\% & 1.5 & 1.0\% & 1.5 & 1.6\% & 1.3 & 2.0\% & 0.8 \\
$\langle K^+|(us)_Ld_L|p\rangle$ & 7.5\% & 1.6 & 2.3\% & 2.2 & 7.2\% & 1.5 & 3.3\% & 2.1 & 1.9\% & 2.7 \\
$\langle K^+|(ud)_Rs_L|p\rangle$ & 1.6\% & 0.9 & 1.0\% & 1.2 & 1.2\% & 1.1 & 1.3\% & 0.8 & 1.3\% & 0.1 \\
$\langle K^+|(ud)_Ls_L|p\rangle$ & 3.9\% & 1.7 & 2.1\% & 2.4 & 3.3\% & 1.6 & 1.4\% & 2.2 & 1.5\% & 1.7 \\
$\langle K^+|(ds)_Ru_L|p\rangle$ & 2.7\% & 1.0 & 2.3\% & 0.8 & 1.4\% & 1.1 & 2.3\% & 1.0 & 0.7\% & 0.8 \\
$\langle K^+|(ds)_Lu_L|p\rangle$ & 2.1\% & 1.8 & 1.5\% & 2.4 & 1.4\% & 1.8 & 0.8\% & 2.2 & 1.6\% & 0.8 \\
\hline
$\langle \eta|(ud)_Ru_L|p\rangle$ & 39.7\% & 1.0 & 31.7\% & 1.0 & 23.8\% & 1.4 & 9.4\% & 1.0 & 4.7\% & 1.6 \\
$\langle \eta|(ud)_Lu_L|p\rangle$ & 2.8\% & 1.0 & 1.3\% & 1.2 & 2.5\% & 1.1 & 1.9\% & 1.8 & 1.5\% & 0.7 \\
\hline\hline
\end{tabular}
\end{center}
\end{table}

\begin{table}
\begin{center}
\caption{Table of renormalized $W_0$ in the physical kinematics at 2 GeV in $\overline{\rm MS}$ NDR scheme. The fourth column presents relative error of systematic uncertainties; ``$\chi$'' is coming from chiral extrapolation given from three different fitting ranges as explained in the text, ``$q^4$'' and ``$a^2$'' column is uncertainty of higher order correction than $\mathcal O(q^2)$ and lattice artifact at $\mathcal O(a^2)$ respectively. ``$m_s$'' column is uncertainty for use of unphysical strange quark mass. $\Delta_Z$ and $\Delta_a$ are error of renormalization factor and lattice scale estimate respectively.
}
\label{tab:w0_error}
\begin{tabular}{r|c|c|c|ccc|cccc}
\hline\hline
\multirow{2}{*}{Matrix element} & \multirow{2}{*}{$W_0$ GeV$^2$} &
 \multirow{2}{*}{stat.[\%]} & \multicolumn{7}{c}{Systematic error [\%]}\\
 & & & total & $\chi$ & $q^4$ & $mq^2$ & $a^2$ & $m_s$ & $\Delta_a$ & $\Delta_Z$\\
\hline
$\langle \pi^0|(ud)_Ru_L|p\rangle$ & -0.131(4)(13) & 3.0 & 9.7 & 1.8 & 0.7 & 0.3 &
\multirow{4}{*}{5.0} & \multirow{4}{*}{-} & \multirow{4}{*}{0.6} & \multirow{4}{*}{8.1}\\
$\langle \pi^0|(ud)_Lu_L|p\rangle$ & 0.134(5)(16) & 3.4 & 11.6 & 5.7 & 2.3 & 2.6 \\
$\langle \pi^+|(du)_Rd_L|p\rangle$ & -0.186(6)(18) & 3.0 & 9.7 & 1.8 & 0.7 & 0.3 \\
$\langle \pi^+|(du)_Ld_L|p\rangle$ & 0.189(6)(22) & 3.4 & 11.6 & 5.7 & 2.3 & 2.6 \\
\hline
$\langle K^0|(us)_Ru_L|p\rangle$ & 0.103(3)(11) & 2.8 & 10.4 & 2.8 & 0.7 & 1.1 &
\multirow{8}{*}{5.0} & \multirow{8}{*}{3.0} & \multirow{8}{*}{0.6} & \multirow{8}{*}{8.1}\\
$\langle K^0|(us)_Lu_L|p\rangle$ & 0.057(2)(6) & 3.5 & 10.7 & 3.1 & 1.0 & 2.1 \\
$\langle K^+|(us)_Rd_L|p\rangle$ & -0.049(2)(5) & 3.7 & 10.9 & 3.5 & 1.6 & 2.0 \\
$\langle K^+|(us)_Ld_L|p\rangle$ & 0.041(2)(5) & 4.4 & 13.1 & 7.5 & 3.3 & 1.9 \\
$\langle K^+|(ud)_Rs_L|p\rangle$ & -0.134(4)(14) & 3.2 & 10.3 & 1.6 & 1.3 & 1.3 \\
$\langle K^+|(ud)_Ls_L|p\rangle$ & 0.139(4)(15) & 3.0 & 10.9 & 3.9 & 1.4 & 1.5 \\
$\langle K^+|(ds)_Ru_L|p\rangle$ & -0.054(2)(6) & 3.6 & 10.6 & 2.7 & 2.3 & 0.7 \\
$\langle K^+|(ds)_Lu_L|p\rangle$ & -0.098(3)(10) & 2.8 & 10.3 & 2.1 & 0.8 & 1.6 \\
\hline
$\langle \eta|(ud)_Ru_L|p\rangle$ & 0.006(2)(3) & 30.0 & 42.1 & 39.7 & 9.4 & 4.7 &
\multirow{2}{*}{5.0} & \multirow{2}{*}{-} & \multirow{2}{*}{0.6} & \multirow{2}{*}{8.1}\\
$\langle \eta|(ud)_Lu_L|p\rangle$ & 0.113(3)(12) & 3.1 & 10.2 & 2.8 & 1.9 & 1.5 \\
\hline\hline
\end{tabular}
\end{center}
\end{table}

\begin{table}
\begin{center}
\caption{Table of renormalized $W_\mu$ (Eq.~\ref{eq:w_mu}),
which is the form factor in the physical kinematics with final state of $\mu^+$.}
\label{tab:wmu}
\begin{tabular}{r|c}
\hline\hline
Matrix element & $W_\mu$ GeV$^2$ \\
\hline
$\langle \pi^0|(ud)_Ru_L|p\rangle$ & -0.118(3)(12)\\
$\langle \pi^0|(ud)_Lu_L|p\rangle$ &  0.119(4)(14)\\
$\langle \pi^-|(du)_Ru_L|n\rangle$ & -0.167(4)(16)\\
$\langle \pi^-|(du)_Lu_L|n\rangle$ & 0.169(5)(20)\\
\hline
$\langle K^0|(us)_Ru_L|p\rangle$ & 0.099(2)(10)\\
$\langle K^0|(us)_Lu_L|p\rangle$ & 0.061(2)(7)\\
\hline
$\langle \eta|(ud)_Ru_L|p\rangle$ & 0.011(2)(3)\\
$\langle \eta|(ud)_Lu_L|p\rangle$ & 0.108(3)(11)\\
\hline\hline
\end{tabular}
\end{center}
\end{table}

\begin{figure}
\begin{center}
\includegraphics[width=120mm]{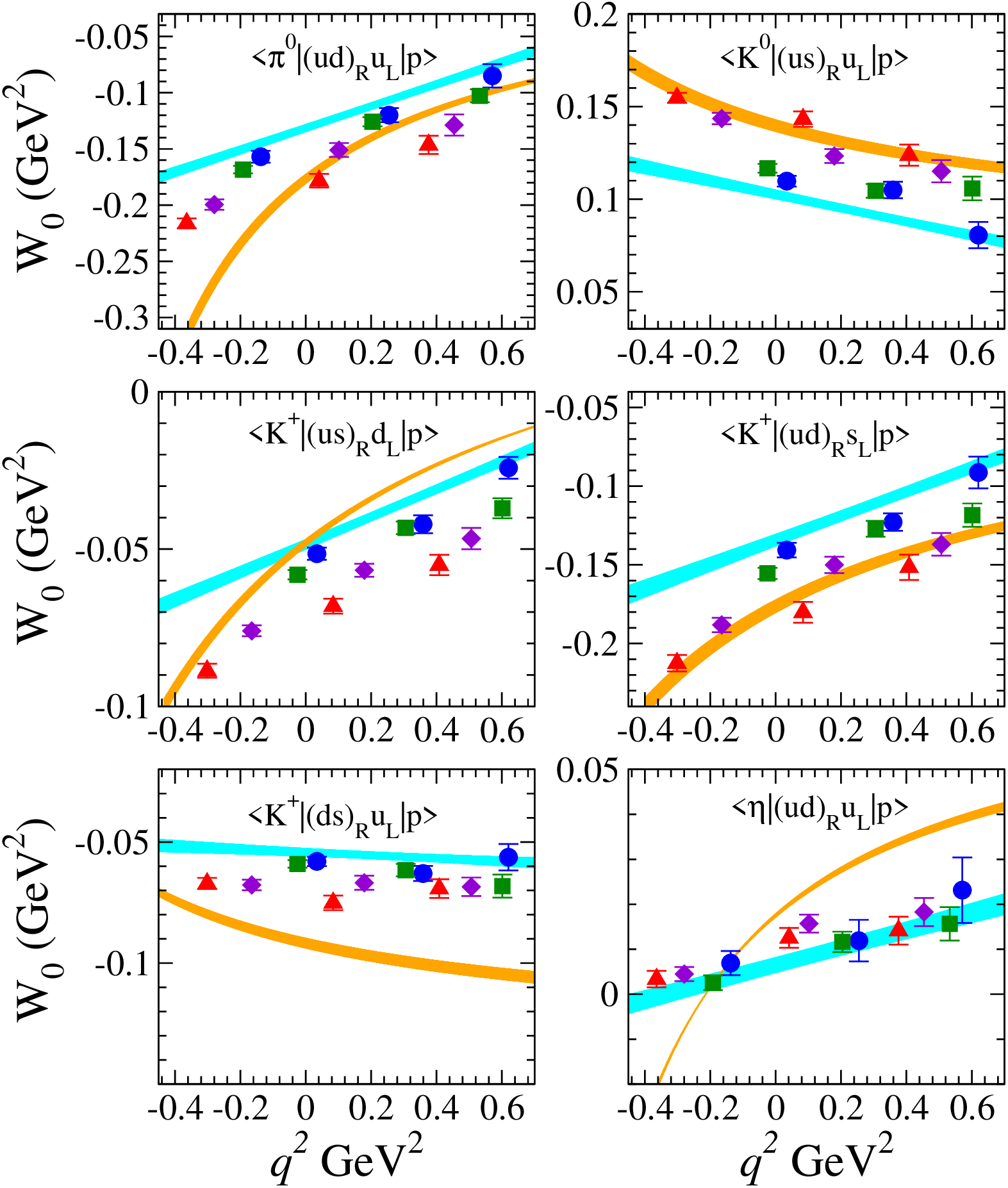}
\vskip 3mm
\caption{
Renormalized $W_0$ with $\overline{\rm MS}$ scheme in $\mu=2$ GeV for $\Gamma=R$ at each channel. The different symbols are for $m=0.005$ (circle), 0.01 (square), 0.02 (diamond) and 0.03 (triangle). We also show the chiral extrapolation line in physical pseudoscalar mass as cyan colored band. Orange colored band show the $W_0^{\alpha,\beta}$ including central value and error of LECs obtained in our calculation. Those error bands only include statistical error.
}
\label{fig:w0_q2dep_R}
\end{center}
\end{figure}

\begin{figure}
\begin{center}
\includegraphics[width=120mm]{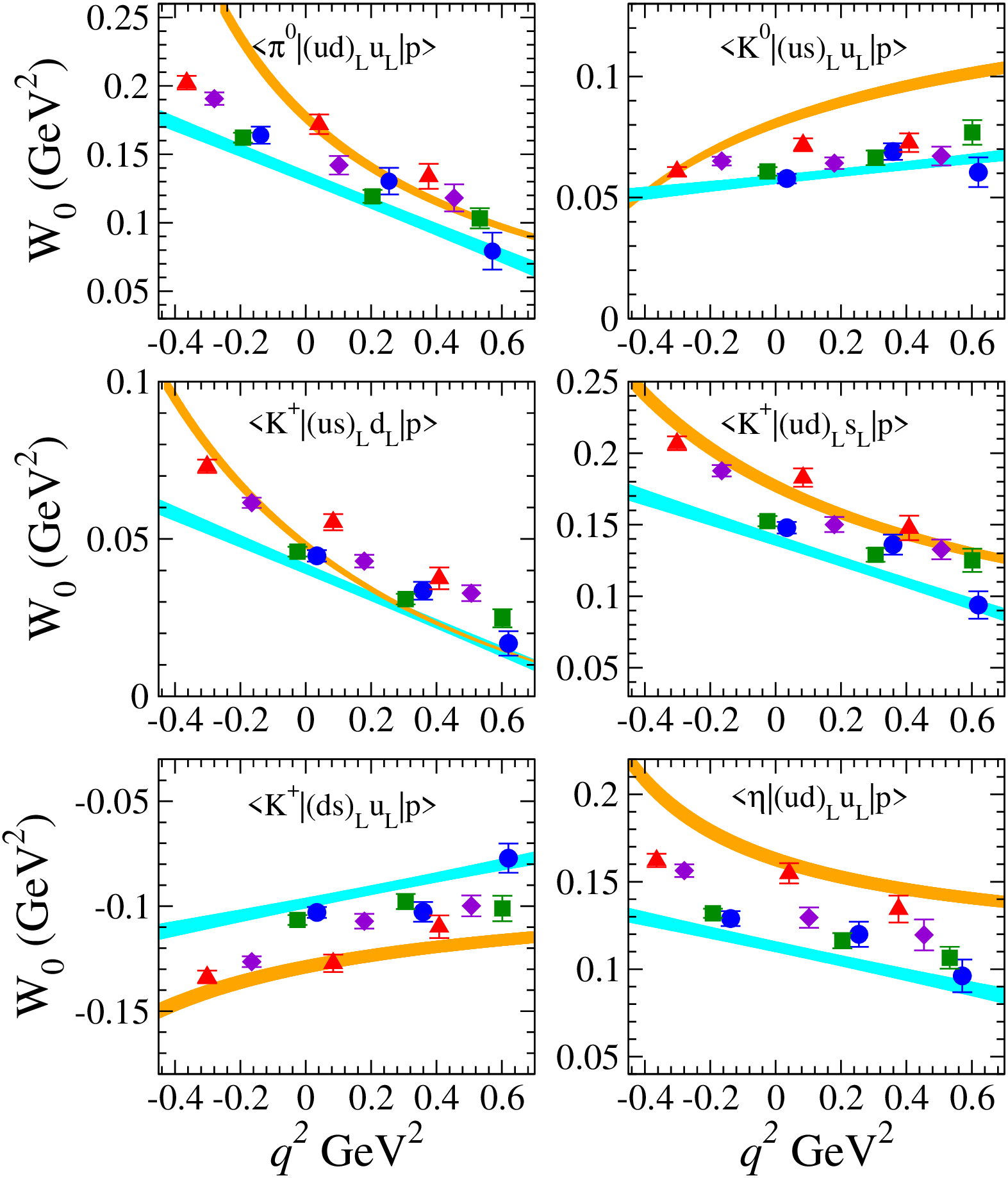}
\vskip 3mm
\caption{
Renormalized $W_0$ with $\overline{\rm MS}$ scheme in $\mu=2$ GeV for $\Gamma=L$. Each symbol is same as Figure \ref{fig:w0_q2dep_R}.
}
\label{fig:w0_q2dep_L}
\end{center}
\end{figure}

%\begin{figure}
%\begin{center}
%  \includegraphics[width=110mm]{W0_fitdep_elem0_indv_R.eps}
%  \vskip 5mm
%  \includegraphics[width=110mm]{W0_fitdep_elem0_indv_L.eps}
%  \vskip 3mm
%  \caption{Plot of a result after performing $q^2=0$ extrpolation using data obtained by plateau fit with three different fitting ranges, $[13,17]$, $[11,17]$ and $[10,18]$ at each quark mass.}
%\label{fig:w0_fitdep_q0}
%\end{center}
%\end{figure}

\begin{figure}
\begin{center}
\includegraphics[width=120mm]{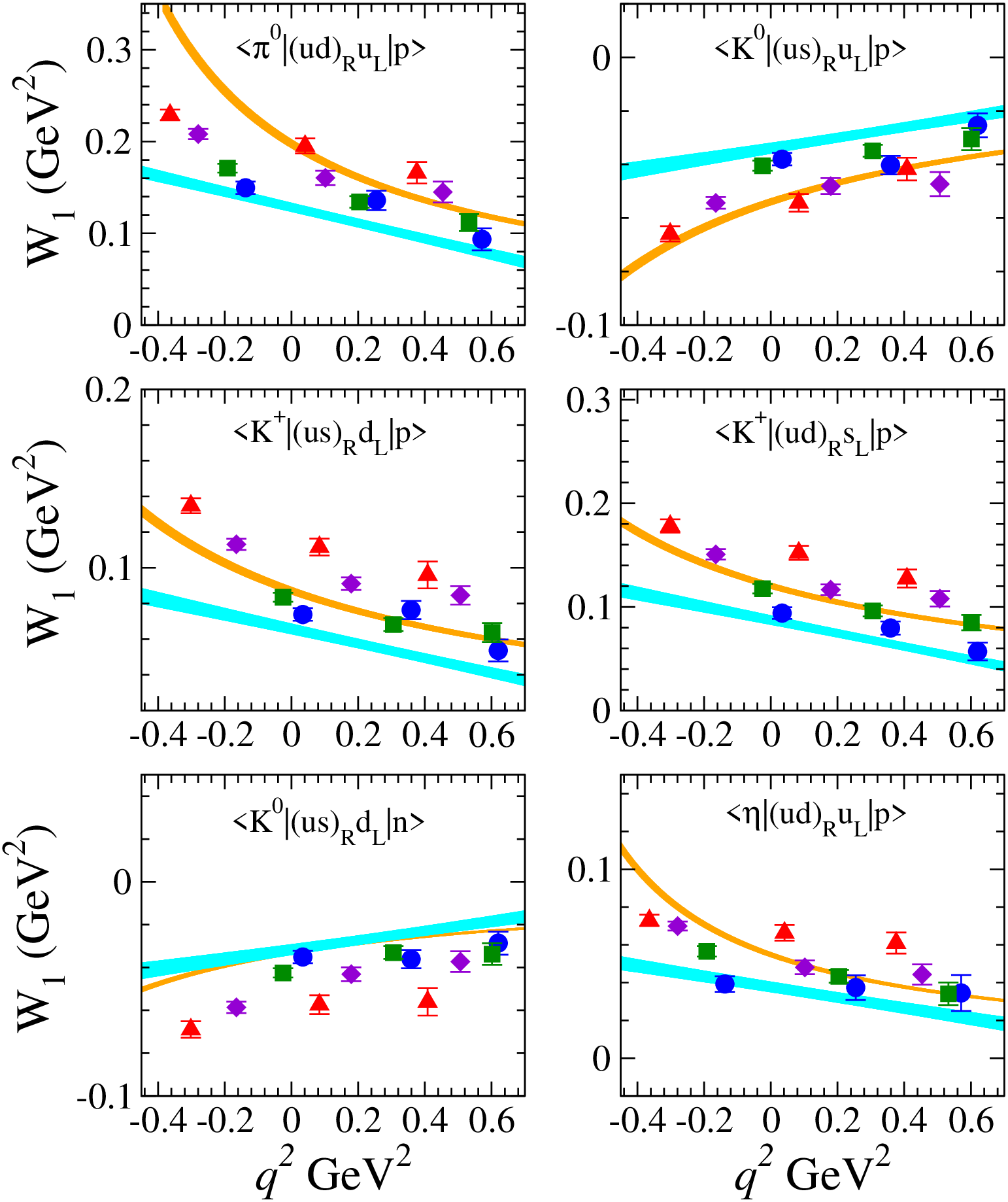}
\vskip 3mm
\caption{
  Renormalized $W_1$ for $\Gamma=R$. Each symbol is same as Figure \ref{fig:w0_q2dep_R}.
}
\label{fig:w1_q2dep_R}
\end{center}
\end{figure}

\begin{figure}
\begin{center}
\includegraphics[width=120mm]{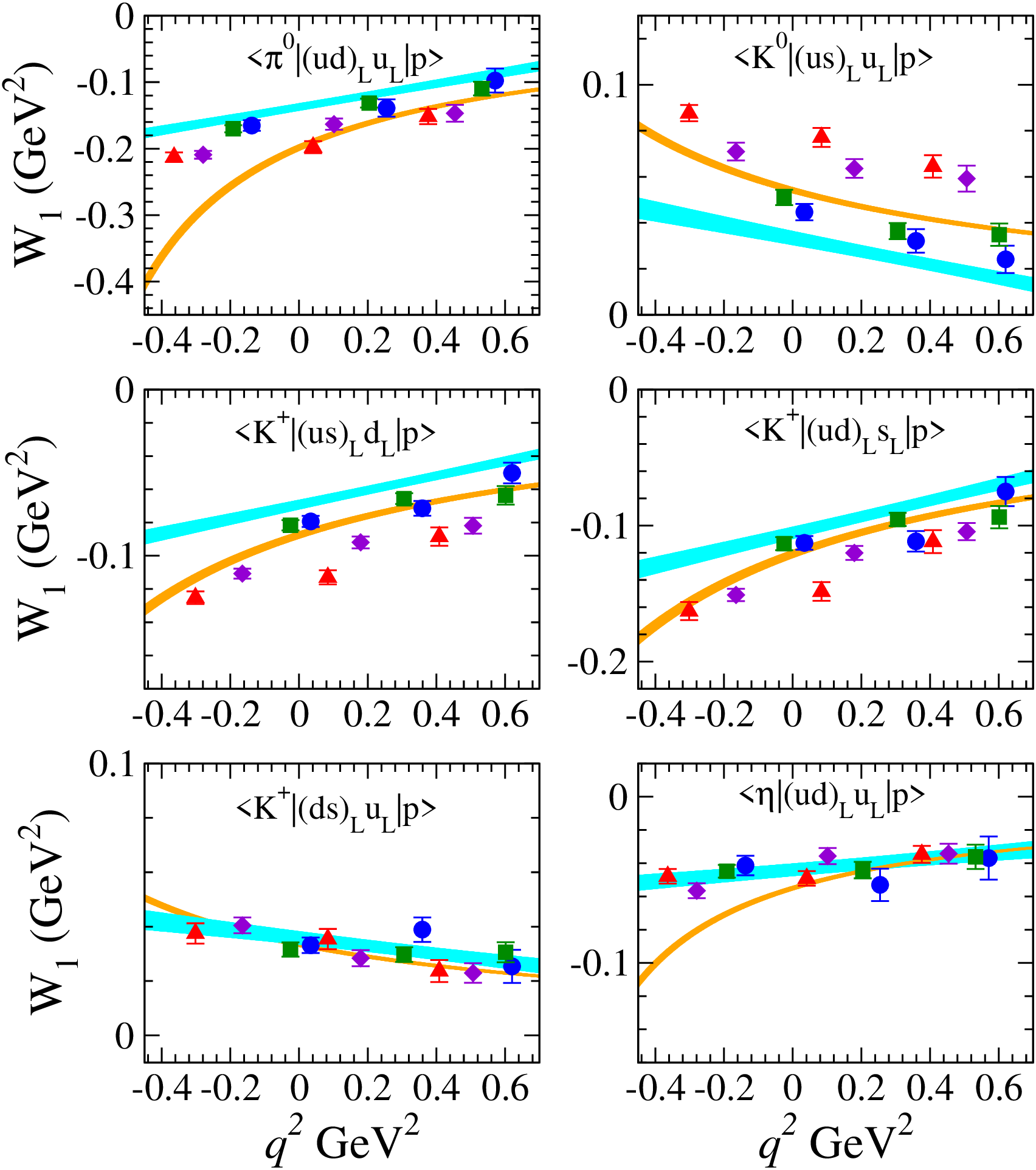}
\vskip 3mm
\caption{
Renormalized $W_1$ for $\Gamma=L$. Each symbol is same as Figure \ref{fig:w0_q2dep_R}.
}
\label{fig:w1_q2dep_L}
\end{center}
\end{figure}

\clearpage
\pagebreak

\section{Application to the kinematics of dark matter model}\label{sec:IND}
In this section, we present a demonstration of the interesting applications to the model using other kinematics, in which energetic pion is emitted from proton and dark matter (DM) appears instead of lepton. According to \cite{Davoudiasl:2010am,Davoudiasl:2011fj,Davoudiasl:2014gfa}, the so-called ``induced nucleon decay (IND)'' scenario, the proton should decay to DM particles, $(\Psi,\Phi)$, having the anti-baryon number with mass $m_{\Phi,\Psi}\sim 2$--3 GeV. This model, motivated by hypothesis of asymmetric DM model \cite{Kaplan:2009ag}, assumes the net baryon number in the Universe is symmetric, in which the SM particle sector has a baryon number asymmetry while the particle $X_1$ in hidden sector has opposite asymmetry, and DM $(\Psi,\Phi)$ has been generated from $X_1$ decay in the early Universe. 
%{\color{red} \it YA: does the statement ``a particle has or do not have
%asymmetry'' make sense ? }
Under a consistency with Sakharov condition, such decay should
have baryon number violation and CP violation in non-thermal
circumstance.  In IND model, nucleon and pseudoscalar are interacting
with DM through $X_1$, and thus scattering process $\Psi N\rightarrow
\Phi^\dag P$ and $\Phi N\rightarrow \bar\Psi P$ occur. The interesting
feature of this model is that the QCD matrix element is same as that
of the standard nucleon decay, since the operator related to DM
scattering is composed of effective three-quark interaction,  
\begin{equation}
  u_Rd_Rd_R\Psi_R\Phi/\Lambda^3 + \textrm{hc},
  \label{eq:leff_ind}
\end{equation}
and only difference is its kinematics of which $q^2$ is different from
on-shell lepton. In principle lattice calculation is accessible to the
matrix element at $q^2$ values relevant to this model, and so that we
can also provide more accurate value for the prediction of this model.  

\begin{figure}
\begin{center}
  \includegraphics[width=120mm]{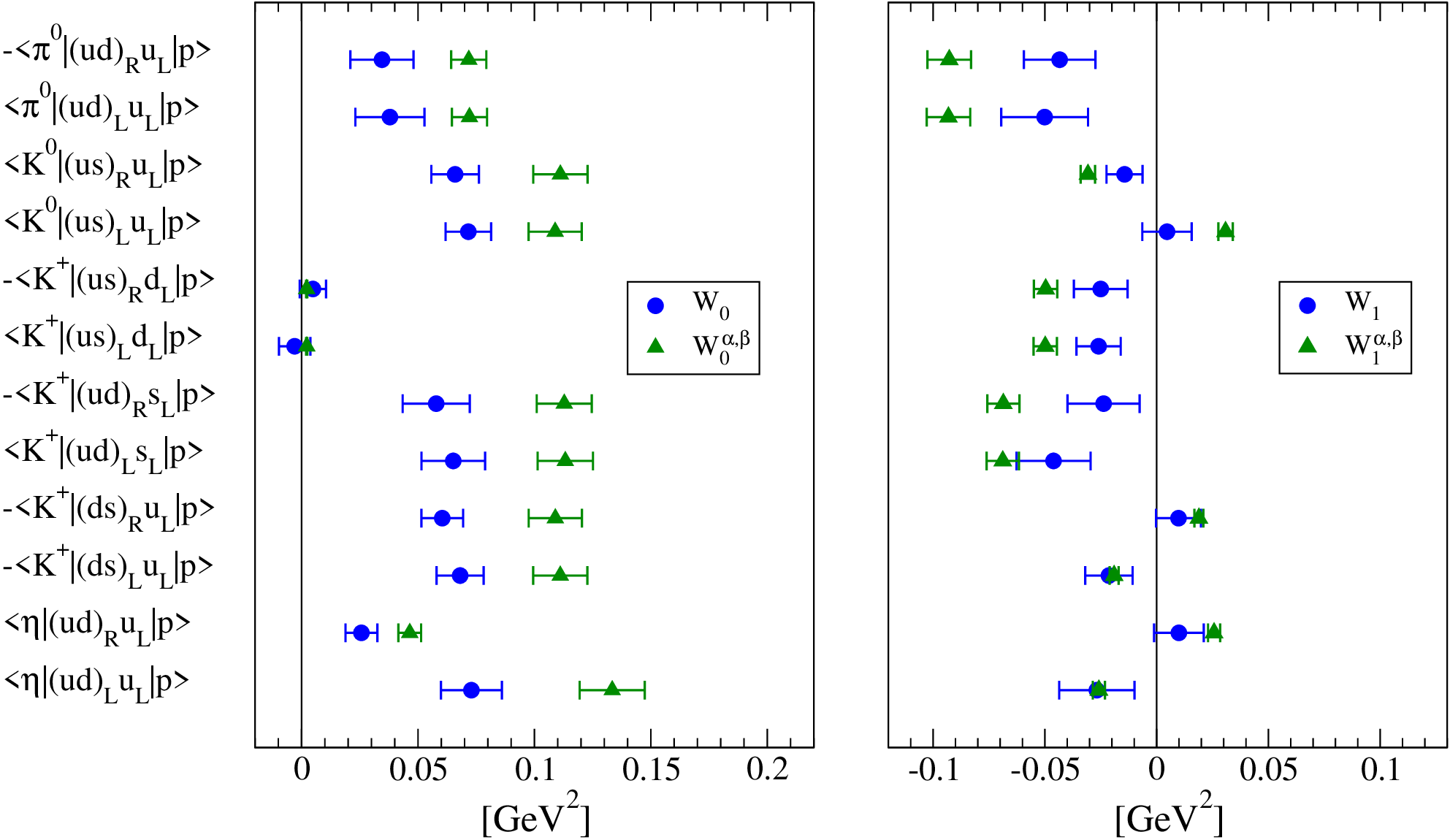}
  \vskip 2mm
\caption{
Summary of matrix elements ``$W_{0,1}$, $W_{0,1}^{\alpha,\beta}$'' in typical meson momentum $|\vec p|=1$ GeV for IND model.
}
\label{fig:w01_ind}
\end{center}
\end{figure}

The DM mass $m_{\Phi,\Psi}\sim$2--3 GeV is predicted from cosmological
observation and DM stability, which is much heavier than lepton mass, so
that under momentum conservation pion has finite momentum, which is a
shifted region to $-q^2<0$, (right direction from zero in
Figure~\ref{fig:w0_q2dep_R}, \ref{fig:w0_q2dep_L}, \ref{fig:w1_q2dep_R}
and \ref{fig:w1_q2dep_L}). Recalling the formula of transition form
factor in Eq.~(\ref{eq:formfactor}), relevant form factor is both $W_0$
and $W_1$, since DM mass is heavy, $q^2\sim 4m^2_{\Phi,\Psi} >
m_p^2$. Typical meson momentum in IND model is $|\vec p|=1$ GeV, in
which the kinematics of IND model is $q^2\simeq1$
GeV$^2$. Figure~\ref{fig:w01_ind} plots $W_{0,1}$ and
$W_{0,1}^{\alpha,\beta}$ extrapolated to $q^2=1$ GeV$^2$ using lattice
results. Focusing on the pion channel, one sees that ``direct'' lattice
calculation provides 25--50\% value of $W_{0,1}^{\alpha,\beta}$ used for
an estimate of proton lifetime in IND model
\cite{Davoudiasl:2010am,Davoudiasl:2011fj,Davoudiasl:2014gfa}. Concerning
the convergence issue of BChPT at tree-level applying to energetic
meson arises in this kinematics, our lattice result indicates such a
difference from an evaluation based on tree-level BChPT may not be
negligible. One sees that possible effect to proton decay amplitude when
using $W_{0,1}$ in our results may be factor 4 and more suppression to
the results obtained with BChPT. This potentially large systematic error
needs to be considered when one use the BChPT for this purpose.

\section{Summary and discussion}\label{sec:summary}
In this paper we present improved computation of proton decay matrix
element using the all-mode-averaging technique on the $N_f=2+1$
domain-wall fermion configurations. Compared to previous work
\cite{Aoki:2013yxa} (also see Table~\ref{tab:history}), the statistical
error has been significantly reduced for both low-energy constant in
baryon chiral perturbation theory (BChPT) and matrix element extracted
from three-point function. Our analysis using the precise lattice data
with three variation of momentum, by which we add one more higher $q^2$,
can evaluate higher order correction than $\mathcal O(q^2)$. The
systematic uncertainty for the chiral extrapolation due to using
unphysical pion around $m_\pi=0.33$ GeV, is still large rather than
$\mathcal O(q^4)$, $\mathcal O(mq^2)$ correction, while its magnitude
strongly depends on the chirality of baryon number violating
operator. This uncertainty can be reduced by using larger volume than 3
fm$^3$ in physical pion mass generated by RBC-UKQCD collaboration
\cite{Blum:2014tka} in future work. Currently the dominated error is
coming from the uncertainty of renormalization factor and lattice
artifact correction, and those may be reduced by the further effort of
renormalization scheme and comparison with finer lattice
\cite{Blum:2014tka}. Final result of $W_{0,\mu}$ is presented in Table
\ref{tab:w0_error} and \ref{tab:wmu}, in which the total error in pion
channel for both $e,\nu$ and $\mu$ final state is 10--14\%, and of kaon
sector is also of similar precision. Compared to
$W_{0,\mu}^{\alpha,\beta}$ via ``indirect'' method using the improved
lattice calculation of LECs, $W_{0,\mu}$ from ``direct'' method is
1.3--1.4 time small for proton decay amplitude. This means, if our
result of $W_{0,\mu}$ is incorporated into GUT model prediction instead
of $W_{0,\mu}^{\alpha,\beta}$, the proton lifetime prediction may become
about 2 times larger. Finally we note that our calculation is also
applicable to the other kinematics corresponding to a dark matter model,
and pointing out that there will be higher order correction than NLO
BChPT. Our lattice calculation of $W_{0,\mu}$ can provide more
reliable value for such a model. 

\begin{table}
\begin{center}
  \caption{Comparison of matrix element calculation in lattice QCD from
 $N_f=0$ to $N_f=3$ with several groups. The errors of $\alpha$, $\beta$
 and $W_0$ are denoted as the total one, which is combined with
 statistical and systematic errors in the quadrature. We remark that
 $^*$ denotes the result using renormalization constants with
 perturbative matching factors having an error as explained in the text.
 With the error corrected the values would increase as much as $\sim$ 7 \%.
 %{\it \color{red} YA: do QCDSF suffer from same problem ?}
 %\red{PUBLISHED YEAR, CORRECTED Z FACTOR FOR RBC/UKQCD?, JLQCD for ALPHA,BETA IN CONTINUUM, QCDSF REFERENCE?}
  }
\label{tab:history}
\begin{tabular}{c|c|c|c|c|c|c}
\hline\hline
%\multirow{2}{*}{Ref.} & Fermion & \multirow{2}{*}{$N_f$} & Volume & \multirow{2}{*}{$a$ (fm)} & $m_\pi$
%& \multirow{2}{*}{Renorm.} \\
%& action & & (fm$^3$) & & (GeV) & \\
%\hline
%\cite{Aoki:1999tw}  & \multirow{2}{*}{Wilson} & \multirow{2}{*}{0} & \multirow{2}{*}{2.4} 
%& \multirow{2}{*}{0.09} & \multirow{2}{*}{0.45--0.73} & One-loop & $\alpha=-0.015(1)$\\
%JLQCD(2000) & & & & & & perturbation & $\beta=0.014(1)$\\
% & & & & & & & $W_0^{LR}(p\rightarrow\pi^0)$\\
% & & & & & & & $=-0.134(16)$\\
Ref. & JLQCD & CP-PACS & RBC  & QCDSF & RBC/ & This \\
     &       & \& JLQCD &     &       & UKQCD & work\\
 & (2000) & (2004) & (2007) & (2008) & (2008,2014) & \\
 & \cite{Aoki:1999tw} & \cite{Tsutsui:2004qc} & \cite{Aoki:2006ib} & \cite{Braun:2008ur} & \cite{Aoki:2008ku,Aoki:2013yxa} & \\
\hline
Fermion & Wilson & Wilson & DW & Wilson & DW & DW\\
\hline
$N_f$ & 0 & 0 & 0 and 2 & 2 & 3 & 3\\
\hline
 & & \multirow{4}{*}{(3.3)$^3$} & Quench & \multirow{4}{*}{(1.68)$^3$} & \multirow{4}{*}{(2.65)$^3$} & \multirow{4}{*}{(2.65)$^3$}\\
Volume & (2.4)$^2$ & & (1.6)$^3$ & & &\\
(fm$^3$) & $\times$4.1 & & Two-flavor & & & \\
 & & & (1.9)$^3$ & & & \\
\hline
\multirow{4}{*}{$a$ (fm)} & \multirow{4}{*}{0.09} & \multirow{4}{*}{0} & Quench & \multirow{4}{*}{0.07} & \multirow{4}{*}{0.11} & \multirow{4}{*}{0.11}\\
 & & & 0.1 & & & \\
 & & & Two-flavor & & & \\
 & & & 0.12 & & & \\
\hline
 & \multirow{4}{*}{0.45--0.73} & \multirow{4}{*}{0.6--1.2} & Quench & \multirow{4}{*}{0.42--1.18} & \multirow{4}{*}{0.34--0.69} & \multirow{4}{*}{0.34--0.69} \\
 $m_\pi$ & & & 0.39--0.58 & & &\\\
 (GeV)& & & Two-flavor & & &\\\
 & & & 0.48--0.67 & & &\\
\hline
Renorm. & One-loop & One-loop & NPR & NPR & NPR & NPR \\
$\mu$ & $1/a$, $\pi/a$ & 2 GeV& 2 GeV & 2 GeV & 2 GeV & 2 GeV\\
\hline
 & \multirow{4}{*}{$-0.015(1)$} & \multirow{4}{*}{$-0.0090(^{+10}_{-21})$} & Quench & \multirow{4}{*}{$-0.0091(4)$}& \multirow{4}{*}{$-0.0119^*(26)$} & \multirow{4}{*}{$-0.0144(15)$}\\
$\alpha$ & & & $-0.0100^*(19)$ & & &\\
(GeV$^3$) & & & Two-flavor & & & \\
 & & & $-0.0118^*(21)$ & & & \\
\hline
 & \multirow{4}{*}{$0.014(1)$} & \multirow{4}{*}{$0.0096(^{+11}_{-22})$} & Quench & \multirow{4}{*}{$0.0090(4)$} & \multirow{4}{*}{$0.0128^*(28)$} & \multirow{4}{*}{$0.0144(15)$}\\
$\beta$ & & & $0.0108^*(21)$ & & &\\
(GeV$^3$) & & & Two-flavor & & & \\
 & & & $0.0118^*(21)$ & & &\\
\hline
$p\rightarrow \pi^0$\\
\hline
 & \multirow{4}{*}{$-0.134(16)$} & \multirow{4}{*}{-} & Quench & \multirow{4}{*}{-} & \multirow{4}{*}{$-0.103^*(41)$} & \multirow{4}{*}{$-0.131(13)$}\\
$W_0^{LR}$ & & & $-0.060^*(18)$ & & &\\
(GeV$^2$) & & & Two-flavor & & & \\
 & & & - & & & \\
\hline
& \multirow{4}{*}{$0.128(17)$} & \multirow{4}{*}{-} & Quench & \multirow{4}{*}{-} & \multirow{4}{*}{$0.133^*(40)$} & \multirow{4}{*}{$0.134(16)$}\\
$W_0^{LL}$  & & & $0.086^*(22)$ & & &\\
(GeV$^2$) & & & Two-flavor & & & \\
 & & & - & & & \\
\hline\hline
\end{tabular}
\end{center}
\end{table}

\section*{Acknowledgments}
We thank members of RIKEN-BNL-Columbia (RBC) and UKQCD collaboration for
sharing USQCD resources for part of our calculation. ES thanks Hooman
Davoudiasl for a useful discussion. Numerical calculations were
performed using the RICC at RIKEN and the Ds cluster at FNAL. This work
was supported by the 
JSPS KAKENHI Grant, Nos.~JP22540301 (TI), JP22224003 (YA), JP16K05320 (YA),
MEXT KAKENHI Grant, Nos.~JP23105714 (ES), JP23105715 (TI),
and U.S. DOE grants DE-SC0012704 (TI and AS).
We are grateful to BNL, the RIKEN BNL
Research Center, RIKEN Advanced Center for Computing and Communication
(ACCC), and USQCD for providing resources necessary for completion of
this work. ES also thanks the INT and organizers of Program INT-15-3
``Intersections of BSM Phenomenology and QCD for New Physics Searches'',
September 14 - October 23, 2015, and Ryuichiro Kitano for his support
from MEXT Grant-in-Aid for Scientific Research on Innovative Areas
(No. JP25105011).  

\appendix
\section{Leading formula of proton decay matrix element in BChPT}
\label{sec:bchpt}
According to BChPT \cite{Claudson:1981gh,Aoki:1999tw}, the relevant matrix element, $W^{\alpha,\beta}_0$, can be represented as
\begin{eqnarray}
  \langle \pi^0|(ud)_{R}u_{L}|p\rangle &=& \frac{\alpha}{\sqrt 2 f}\Big(1+D+F\Big),\\
  \langle \pi^0|(ud)_{L}u_{L}|p\rangle &=& \frac{\beta}{\sqrt 2 f}\Big(1+D+F\Big),\\
  \langle K^0|(us)_{R}u_{L}|p\rangle &=& 
  -\frac{\alpha}{f}\Big(1+(D-F)\frac{m_N}{m_B}\Big),\\
  \langle K^0|(us)_{L}u_{L}|p\rangle &=& 
  \frac{\beta}{f}\Big(1-(D-F)\frac{m_N}{m_B}\Big),\\
  \langle K^+|(us)_{R}d_{L}|p\rangle &=& 
  \frac{\alpha}{f}\frac{2D}{3}\frac{m_N}{m_B},\\
  \langle K^+|(us)_{L}d_{L}|p\rangle &=& 
  \frac{\beta}{f}\frac{2D}{3}\frac{m_N}{m_B},\\
  \langle K^+|(ud)_{R}s_{L}|p\rangle &=&
  \frac{\alpha}{f}\Big(1+\Big(\frac D 3 + F)\frac{m_N}{m_B}\Big),\\
  \langle K^+|(ud)_{L}s_{L}|p\rangle &=&
  \frac{\beta}{f}\Big(1+\Big(\frac D 3 + F)\frac{m_N}{m_B}\Big),\\
  \langle K^+|(ds)_{R}u_{L}|p\rangle &=&
  \frac{\alpha}{f}\Big(1+\Big(\frac D 3 - F)\frac{m_N}{m_B}\Big),\\
  \langle K^+|(ds)_{L}u_{L}|p\rangle &=&
  -\frac{\beta}{f}\Big(1-\Big(\frac D 3 - F)\frac{m_N}{m_B}\Big),\\
  \langle \eta|(ud)_{R}u_{L}|p\rangle &=&
  -\frac{\alpha}{\sqrt 6 f}\Big(1+D-3F\Big),\\
  \langle \eta|(ud)_{L}u_{L}|p\rangle &=&
  \frac{\beta}{\sqrt 6 f}\Big(3-D+3F\Big),
\end{eqnarray}
with low-energy parameters $D=0.80$, $F=0.47$. In this paper, we use $f=0.131$ GeV, $m_N=0.94$ GeV and $m_B=1.15$ GeV \cite{Aoki:2008ku}. 

\section{Test of soft-pion theorem}\label{appx:soft}
%\red{DISCUSSION ON PION ONLY}
In this section, we present the analysis of matrix element in the soft-pion limit. In this limit, each matrix element is described in term of the leading order of BChPT. In order to test the lattice calculation can make a consistent value with BChPT in the soft-pion limit, we calculate matrix element with two ways; one is BChPT using LECs $\alpha$ and $\beta$ obtained by ``indirect'' lattice calculation and the second is matrix element obtained by ``direct'' lattice calculation. Using LECs the matrix element is 
\begin{eqnarray}
  \langle \pi^0|(ud)_Ru_L|p\rangle_{sp} = \frac{\alpha}{\sqrt{2} f_0}P_Lu_N,\quad 
  \langle \pi^0|(ud)_Lu_L|p\rangle_{sp} = \frac{\beta}{\sqrt{2}f_0}P_Lu_N,
\end{eqnarray}
with subscription $sp$ denoting the soft-pion limit, which corresponds to $p_\mu\rightarrow0$ and chiral limit. On the other hand, the left-hand-side of the above equation is also represented as, 
\begin{eqnarray}
  \langle \pi^0|(ud)_\Gamma u_L|p\rangle_{sp} = P_L W_{sp}u_N, 
\end{eqnarray}
in which $W_{sp}$ is obtained from the form factor at $\vec p=(0,0,0)$ for Eq.~(\ref{eq:formfactor}) in the chiral limit. We define such a form factor as   
\begin{equation}
  W_{\vec p=0} = \lim_{t_1-t,t-t_0\rightarrow \infty} R^\Gamma_3(t,t_1,t_0;0,P_4), 
\end{equation}
and taking the extrapolation into zero quark mass with the linear ansatz, 
\begin{equation}
  W_{\vec p=0} = W_{sp} + c_1 \tilde m. 
  \label{eq:fit_soft}
\end{equation}
Linear ansatz is under the assumption of negligibly small $\sqrt{\tilde m}\sim m_\pi$ term even in $\tilde m\simeq 0$. 

\begin{figure}
\begin{center}
  \includegraphics[width=70mm]{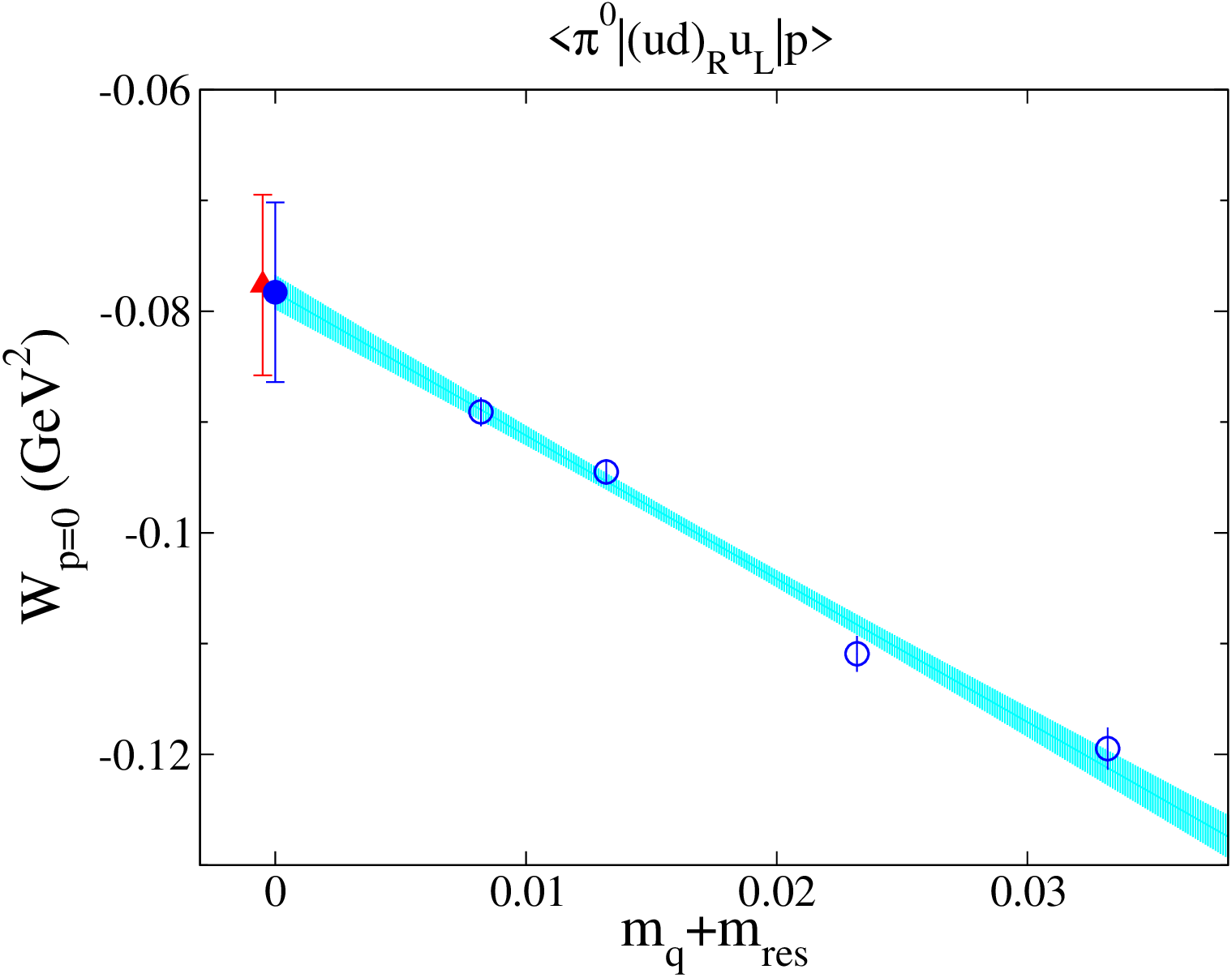}
  \hspace{5mm}
  \includegraphics[width=70mm]{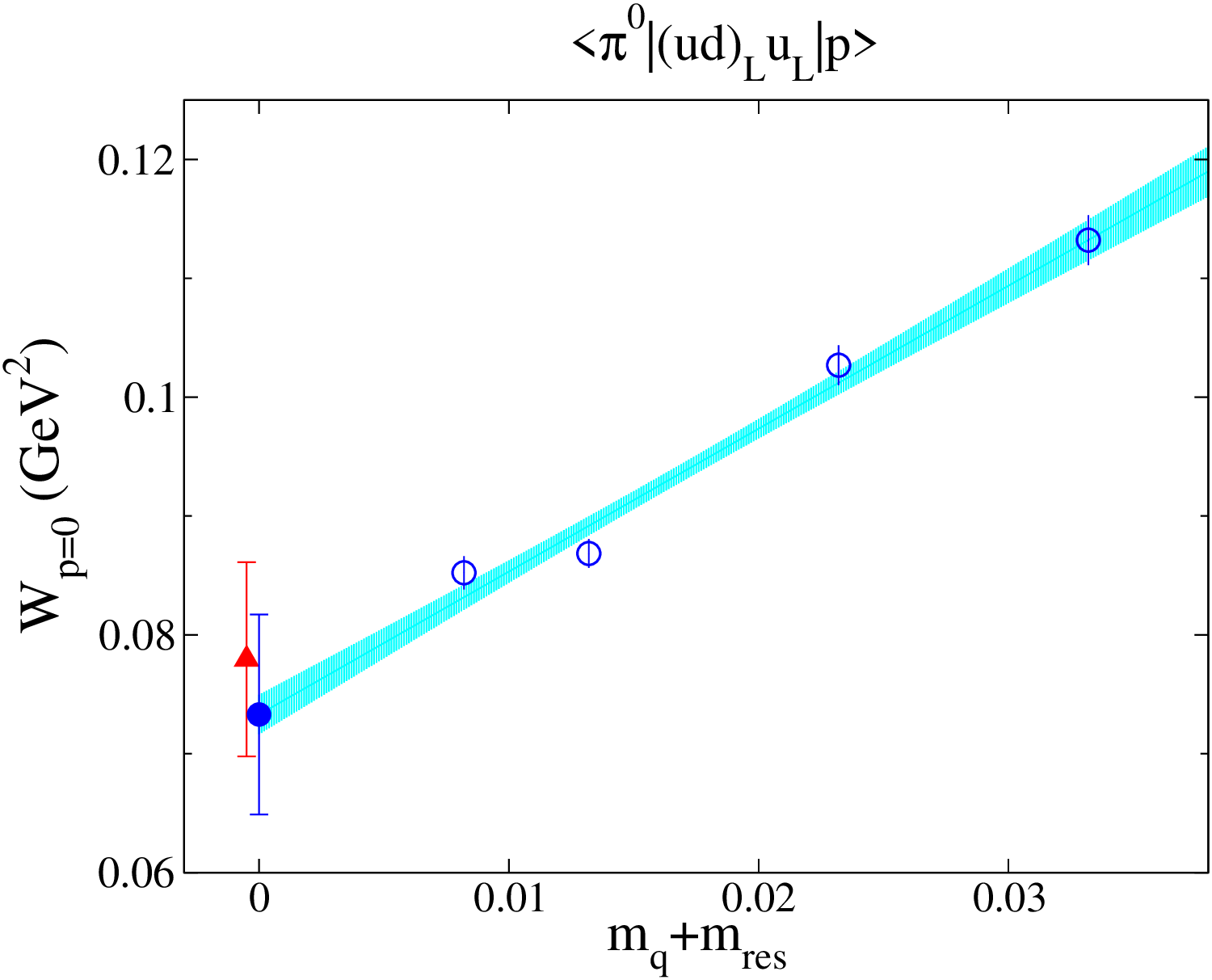}
  \caption{The open circles denote lattice result of $W_{\vec p=0}$ for $RL$ (left) and $LL$ (right) chirality in each quark mass, and colored band shows the fitting function with error. Those errors are the statistical one. In the chiral limit, the triangle and filled circle denote the value of BChPT and extrapolated result in the soft pion limit respectively. Those error bars denote the total error including systematic one, which is obtained by the same procedure in section \ref{sec:ff}.
  }
\label{fig:w_softpi}
\end{center}
\end{figure}

Figure \ref{fig:w_softpi} shows the lattice result of $W_{\vec p=0}$ and $W_{sp}$. We also show the lines of chiral extrapolation and the extrapolated values with linear ansatz in the chiral limit. Here we estimate the systematic uncertainties due to chiral extrapolation by comparing the ``light'' and ``heavy'' region as well as in Table~\ref{tab:w0_chiral}. We also add the uncertainties of renormalization factor and lattice artifact same as in Table~\ref{tab:w0_error}. One sees that the lattice data is close to linear function and the extrapolated value is consistent with BChPT within 1 sigma error. We notice that there is no visible curvature as the square-root of quark mass. It indicates that a coefficient of square-root of quark mass may not be significantly large.

%\red{REFERENCE: ARXIV NUMBER}

%\bibliographystyle{hunsrt}
%\bibliographystyle{h-elsevier}
\bibliographystyle{apsrev4-1}
\bibliography{ref}
\end{document}